\documentclass[twocolumn,epjc3]{svjour3}          

\RequirePackage[T1]{fontenc}

\smartqed  

\RequirePackage{graphicx}
\RequirePackage{subfigure}
\RequirePackage{mathptmx}      
\RequirePackage{flushend}
\RequirePackage[numbers,sort&compress]{natbib}
\RequirePackage[colorlinks,citecolor=blue,urlcolor=blue,linkcolor=blue]{hyperref}

\newcommand\eq[1]  {\begin{eqnarray}
	 #1\end{eqnarray}}
\newcommand\nn  \nonumber

\begin{document}

\title{Low and High Energy Asymptotic Behavior of Electroweak Corrections 
	in Polarized $e^- e^+ \to \mu^- \mu^+$ Process}

\author{A.G.~Aleksejevs\thanksref{e1,addr1}
        \and
        S.G.~Barkanova\thanksref{e2,addr1}
        \and
        Yu.M.~Bystritskiy\thanksref{e3,addr3}
        \and
        V.A.~Zykunov\thanksref{e4,addr3}
}

\thankstext{e1}{e-mail: aaleksejevs@grenfell.mun.ca}
\thankstext{e2}{e-mail: sbarkanova@grenfell.mun.ca}
\thankstext{e3}{e-mail: bystr@theor.jinr.ru}
\thankstext{e4}{e-mail: vladimir.zykunov@cern.ch}

\institute{Memorial University, Corner Brook, Canada\label{addr1}
          \and
           JINR, Dubna, Moscow region, Russia\label{addr3}
}

\date{}

\maketitle

\begin{abstract}
Electroweak radiative corrections will play a major role in the analysis of several upcoming 
ultra-precision experiments such as Belle-II, so is crucial to make sure that they are fully 
under control. The article outlines the recent developments in the theoretical and computational 
app\-ro\-aches to one-loop (NLO) electroweak radiative corrections to the parity-violating asymmetry 
in $e^- e^+ \to \mu^- \mu^+ (\gamma)$ process with longitudinally polarized elec\-trons. 
We derive asymptotic expressions for low and high energy regions (well below or above $Z$-resonance, 
correspondingly) and analyze the leading contributions. For most of energy regions, our results are 
in good agreement with precise computer-algebra based calculation and can used as a quicker alternative.
\end{abstract}

\section{Introduction}
Although the Standard Model of Particle Physics has been extremely successful for several decades, we know it is incomplete, and there has been a lot of excitement generated recently by new physics searchers in both experimental and theoretical communities. 
There are three major ways to search for new physics: with high-energy colliders like LHC (energy frontier), with underground experiments and ground and space-based telescopes (cosmic frontier), or with low-energy but intense particle beams (precision frontier), such as Belle-II or MOLLER. At the precision frontier, the new-generation experiments will be looking for a small, but potentially detectable difference from Standard-Model predictions for decay rates, cross sections and asymmetries, and involve a significant number of Canadian experimentalists and theorists \cite{LRP}.  However, as these experiments become more and more precise and thus challenging, so does the theory input they require.

In this paper, we discuss the one-loop (next-to-the-leading order) electroweak radiative corrections to the parity-violating left-right 
asymmetry in $e^- e^+ \rightarrow \mu^- \mu^+$ process with longitudinally polarized electrons.

Electroweak radiative corrections (EWC) to the electron-positron annihilation have already attracted the significant theoretical attention,
starting from \cite{BH82} where EWC are calculated with arbitrary polarization.
For LEP and SLC colliders, the 4-fermion process  $e^- e^+ \rightarrow f^- f^+$
required the EWC at $Z$-boson pole evaluated with new precision, which was done by collaborations
BHM and WOH \cite{hollik,BH84},
LEPTOP \cite{LEPTOP},
TOPAZ0 \cite{TOPAZ96},
and ZFITTER \cite{ZF91,grup-bar2}.
The ``post-LEP/SLC era''  is provided by  KK \cite{KK} and SANC \cite{sanc-eeff} codes. 

 Recently, program packages such as FeynArts \cite{Hahn:2000kx}, FormCalc \cite{Hahn:1998yk}, LoopTools
\cite{Hahn:1998yk} and FORM \cite{Vermaseren:2000nd}, have created an option of calculating parity-violating NLO effects including all of the possible loop contributions within a given model \cite{Aleksejevs:2007pd}.

The unique feature of our approach is to combine two distinct but mutually reinforcing techniques: semiautomatic, precise, with FeynArts and FormCalc as base languages, and, independently, on paper, with low- and high-energy approximations. Both techniques have their advantages and limitations, but can be very powerful in a combination. Our earlier publications (\cite{Aleksejevs:2010nf}, \cite{Aleksejevs:2012zz}, \cite{Aleksejevs:2010ub}) 
on ($e^- e^- \rightarrow e^- e^-$ scattering showed that the exact analytical one-loop calculations using the computer algebra approach not only increased the theoretical precision dramatically, but also gave us an opportunity to verify previous calculations done in various formalisms.

Basically, we perform the same EWC calculations in two different ways, thus making sure that our evaluations are error-free. Although quite labor-intensive, we suggest that this is the best approach for the analysis of several upcoming ultra-precision experiments with 4-fermion processes such as Belle-II.


Obviously, calculating large sets of one-loop Feynman diagrams on paper is a tedious task.
The packages such as FeynArts \cite{Hahn:2000kx}, FormCalc \cite{Hahn:1998yk}, LoopTools
\cite{Hahn:1998yk} and FORM \cite{Vermaseren:2000nd} allow us to handle the substantial number of
diagrams reasonably quickly, minimize probability of human errors, and avoid the rapid
error accumulation often unavoidable with purely numerical methods.
The one of the key features of the presented work is to compare the complete one-loop set
of electroweak radiative corrections to the parity-violating asymmetry 
in $e^- e^+ \rightarrow \mu^- \mu^+ (\gamma)$ process calculated first on paper, with some approximation, and then within a computational model based on FeynArts, FormCalc and LoopTools, precisely.

FeynArts is a Mathematica package which provides the generation and visualization of Feynman
diagrams and amplitudes involving Standard Model particles. FormCalc, a Mathematica package which reads diagrams generated with FeynArts and evaluates amplitudes with the help of the program FORM in analytical form. LoopTools provides the many-point tensor coefficient functions and is used to numerically evaluate one-loop integrals.
After that, one may implement one of the two renormalization schemes (RS), on-shell or
the constrained differential renormalization (CDR) which is equivalent to $\rm \overline{MS}$ scheme at the one-loop level \cite{Hahn:1998yk}.

Our computation model is not a "black box" and still requires considerable human input on many stages. On the other hand, we can modify these packages to better suit specific projects. In \cite{Aleksejevs:2007pd}, for example,
we adopted FeynArts and FormCalc for the NLO calculations of the differential cross section
in electron-nucleon scattering. In \cite{Aleksejevs:2016whx}, we evaluate higher-order
electroweak effects needed for the accurate interpretation of MOLLER and Belle II experimental
data and show how new-physics particles may enter at the one-loop level.
In general, the results obtained with these packages can be presented in both analytical and
numerical form. Unfortunately, our equations for asymmetry at NLO level obtained with FeynArts and FormCalc are too lengthy and cumbersome to publish. It is also a challenge
putting them into a Monte Carlo as required by the specific experimental analysis.

As we show in the earlier sections, at the certain kinematic conditions, the approximate equations
obtained on paper are in a very good the agreement with the numerical results obtained with computer algebra, and may be used for physical analysis and quick estimations not requiring ultra precision.

\section{Four-fermion process description}

\begin{figure}
	\begin{minipage}{\columnwidth}
		\centering
		\subfigure[]{\includegraphics[width=0.48\textwidth]{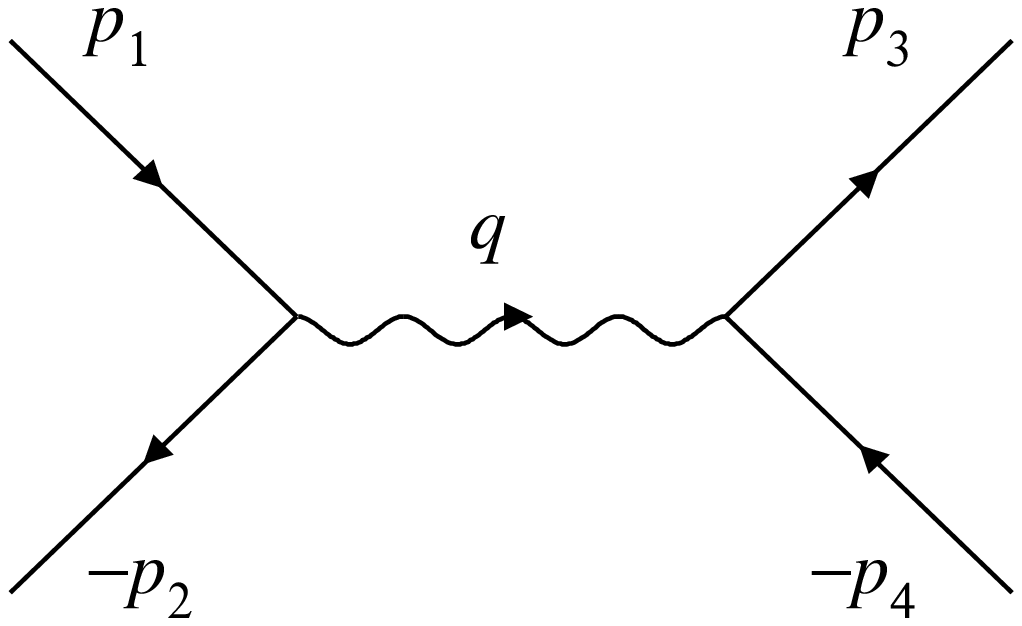}\label{fig.Born}}
		\hfill
		\subfigure[]{\includegraphics[width=0.48\textwidth]{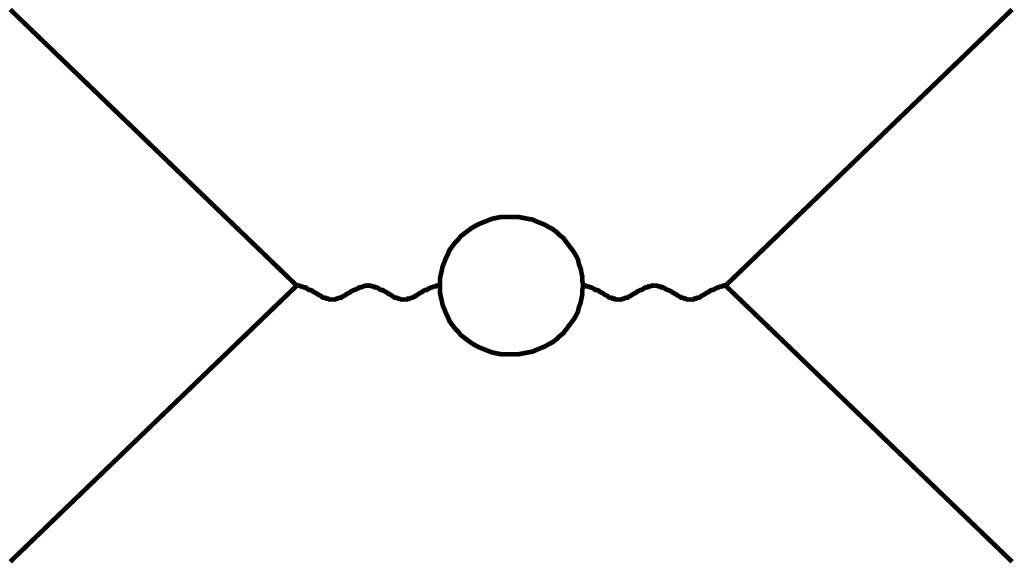}\label{fig.BSE}}
		\\
		\subfigure[]{\includegraphics[width=0.48\textwidth]{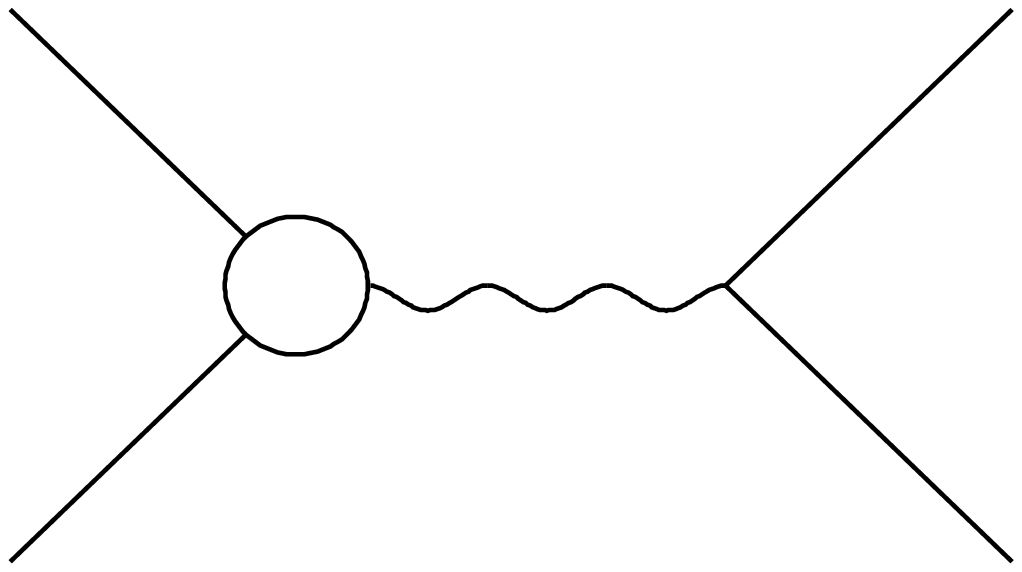}\label{fig.V1}}
		\hfill
		\subfigure[]{\includegraphics[width=0.48\textwidth]{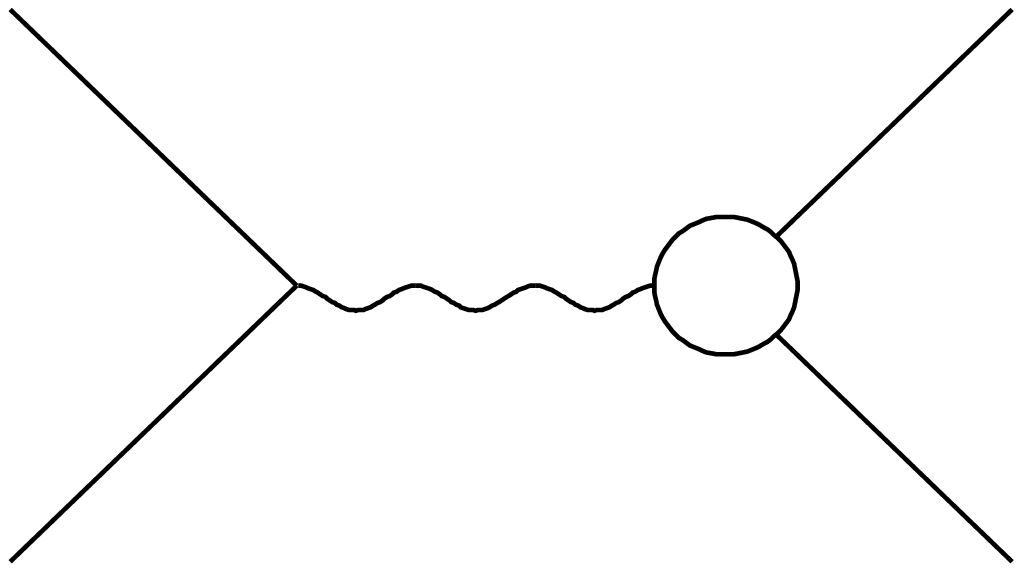}\label{fig.V2}}
		\\
		\subfigure[]{\includegraphics[width=0.48\textwidth]{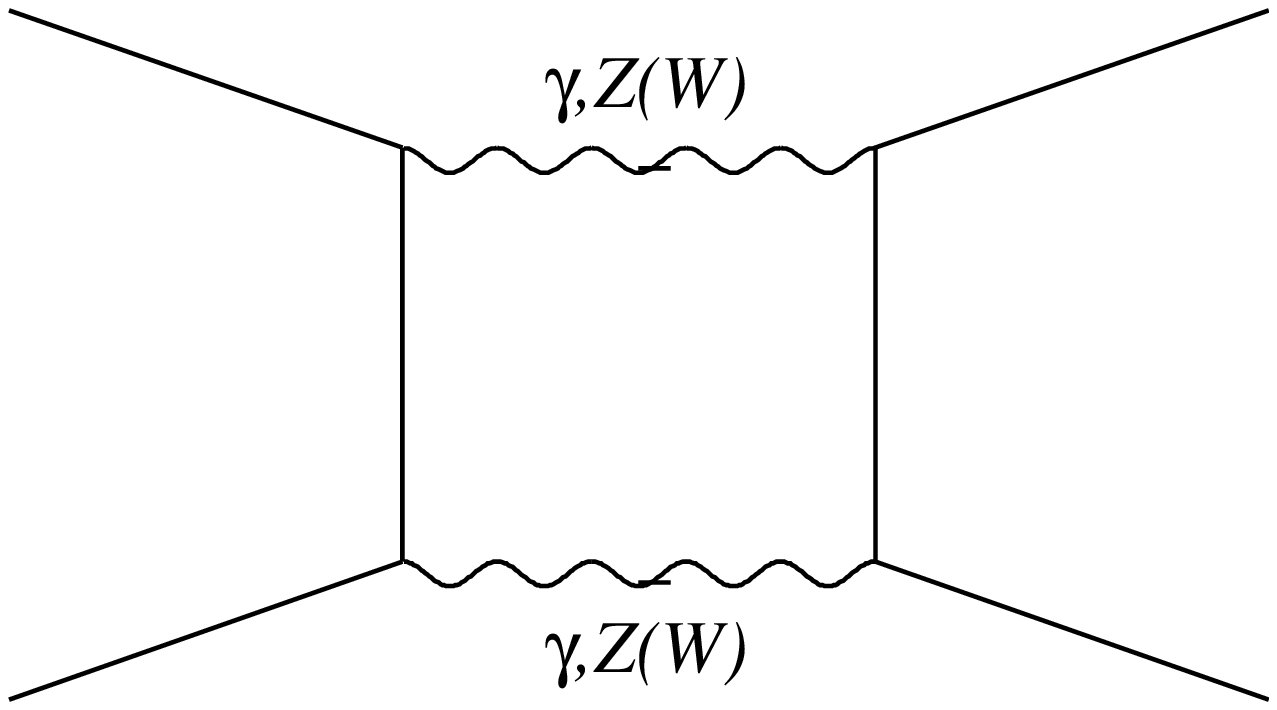}\label{fig.BoxDirect}}
		\hfill
		\subfigure[]{\includegraphics[width=0.48\textwidth]{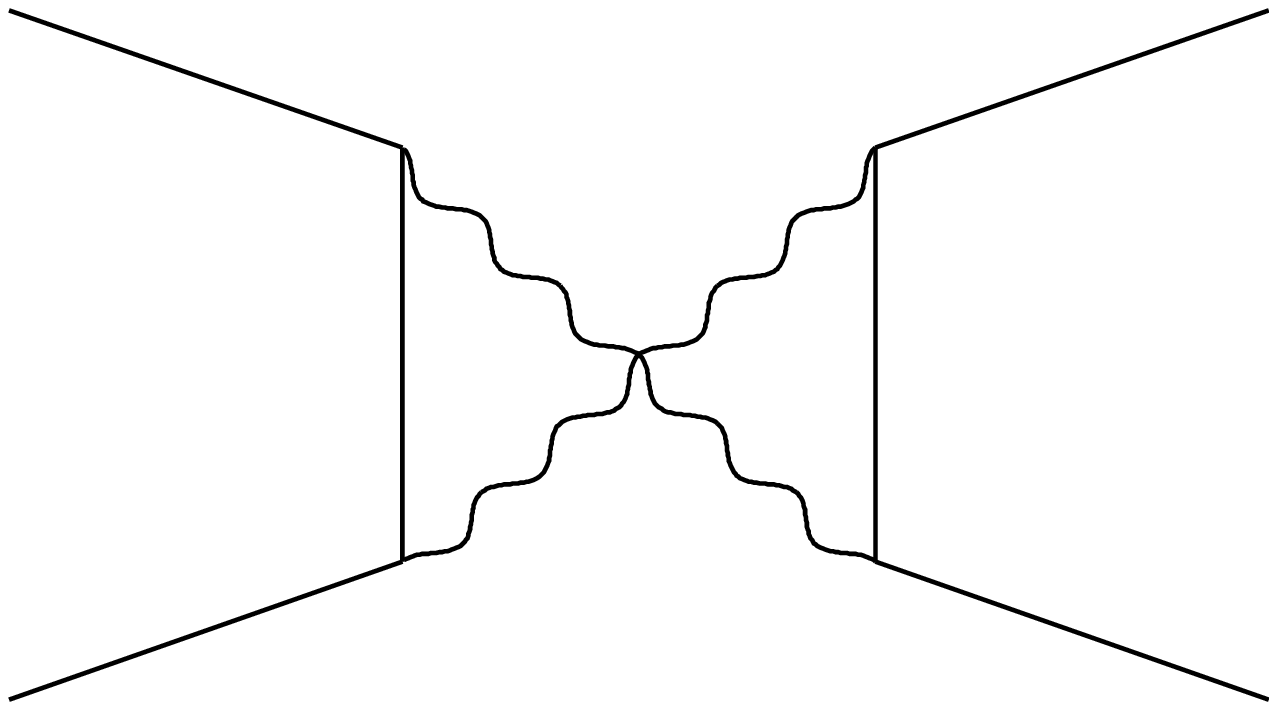}\label{fig.BoxCrossed}}
	\end{minipage}
	\caption{Feynman diagrams of the process 
		$e^- e^+ \rightarrow \mu^- \mu^+$ 
		in radiation-free kinematics:
		\ref{fig.Born} -- Born approximation,
		\ref{fig.BSE} -- boson self energies,
		\ref{fig.V1}, \ref{fig.V2} -- vertex diagrams,
		\ref{fig.BoxDirect}, \ref{fig.BoxCrossed} -- box diagrams.
		Curly lines with no marks denote photon or $Z$-boson.}
	\label{non-rad}
\end{figure}

Let us consider the four-fermion scattering in $s$-channel.
Here we concentrate on the scattering of longitudinally polarized electron off
the unpolarized positron in $s$-channel:
\eq{
	e^-(p_1) + e^+(p_2) \rightarrow \gamma, Z \rightarrow \mu^-(p_3)+\mu^+(p_4).
	\label{0}
}
Feynman graphs for the process (\ref{0}) in tree-level (Born)
and one-loop approximation (NLO)
are presented in Fig.~\ref{non-rad}.

Four-momenta of initial ($p_1$ and $p_2$) and final particles
($p_3$ and $p_4$) form a standard set of Mandelstam invariants
$r$ ($r=s,t,u$):
\eq{
	s=(p_1+p_2)^2,\ t=(p_1-p_3)^2,\ u=(p_2-p_3)^2.
	\label{stu}
}
Unless stated otherwise, we give only ultra-relativistic 
analytical results,
which correspond to the approximation $m_g^2 \ll |r|$.
We use index $g$ for initial and final fermions flavors, i.e.,
in our case, $g=e, \mu$ then $m_e$ is the electron mass and $m_\mu$ is the muon mass.
For the truncated propagator in $s$-channel, we use the following:
\eq{
	D^{j}=\frac{1}{s-m_j^2+im_j\Gamma_j}\ \ (j=\gamma, Z),
	\label{structure}
}
which is present in all amplitudes of Fig.~\ref{non-rad} and
depends on the total energy of the reaction $\sqrt{s}$ in the center-of-mass system (c.m.s.),
intermediate boson mass and its width.
Photon mass $m_\gamma \equiv \lambda $ is equal to zero everywhere except for special cases
mentioned below. In these cases, it is used as an infinitesimal parameter which regularizes 
infrared (IR) divergence.
Mass of $Z$-boson is denoted as $m_Z$, its width is $\Gamma_Z$ (we use scheme with the fixed decay width).

For the differential cross section, we use shortcut notation $\sigma \equiv {d\sigma}/{dc}$, 
where $c=\cos\theta$ and $\theta$ is the angle between initial electron and
final muon detected in c.m.s.
Including one loop, this differential cross section has the form:
\eq{
	\sigma = \frac{\pi^3}{2s} |M_0+M_1|^2 \approx \frac{\pi^3}{2s} (M_0M_0^+ + 2 {\rm Re} M_1M_0^+ ).
	\label{01}
}
The explicit form of Born ($M_0$) and one-loop ($M_1$) amplitudes can be found in \cite{Aleksejevs:2016tjd}.
One loop amplitude $M_1$ has the order of magnitude of ${\cal O}(\alpha^2)$
and consists of boson self energies (BSE), vertexes (Ver) and box diagrams contributions (see Fig.~\ref{non-rad}):
\eq{
	M_{1}=M_{{\rm BSE}}+M_{{\rm Ver}}+M_{{\rm Box}}.
}
In this work, we use the on-shell renormalization scheme \cite{Bohm:1986rj, Denner:1991kt}
with Hollik's renormalization conditions \cite{Bohm:1986rj}.
Thus, the electron self energies are absent.

Born amplitude modulus squared $|M_0|^2$ give raise to Born cross section:
\eq{
	\sigma^0 =\frac{\pi^3}{2s} M_0M_0^+ 
	= \frac{\pi \alpha^2}{s}
	\sum_{i,k=\gamma,Z} D^{i}{D^{k}}^* \mu^{ikik},
	\label{cs0}
}
where the combination
\eq{
	\mu^{ikjl} 
	&=&  b_-^{ikjl} \cdot t^2  +  b_+^{ikjl} \cdot u^2
	\nn\\
	&=&  \lambda_+^{ikjl} \cdot (t^2+u^2)  -  \lambda_-^{ikjl} \cdot (t^2-u^2)
	\label{MU}
}
is identical to one in \cite{Aleksejevs:2016tjd} but has explicitly extracted combinations $b_-$ and $b_+$ 
\eq{
	b_\pm^{ikjl}=\lambda_+^{ikjl} \pm  \lambda_-^{ikjl}
	\label{b}
}
in front of braces with the invariants $t$ and $u$, which is more convenient for analysis.
The combinations $\lambda_\pm^{ikjl}$ can be expressed via the electron polarization degrees $p_B$:
\eq{
	\lambda_+^{ikjl} &=& (\lambda_{e V}^{ik}-p_B \lambda_{e A}^{ik}) \lambda_{\mu V}^{jl},\nn\\
	\lambda_-^{ikjl} &=& (\lambda_{e A}^{ik}-p_B \lambda_{e V}^{ik}) \lambda_{\mu A}^{jl}.
	\label{Lpm}
}
Vector and axial constants of coupling of particle $g$ with photon and $Z$-boson,
\eq{
	v_g^{\gamma}&=&-Q_g, \qquad a_g^{\gamma}=0, \nn\\
	v_g^Z&=&\frac{I_g^3-2Q_gs_{W}^2}{2s_{W}c_{W}}, \qquad
	a_g^Z=\frac{I_g^3}{2s_{W}c_{W}},\nn
}
are combined in the following way:
\eq{
	\lambda_{gV}^{ik}=v_g^iv_g^k + a_g^ia_g^k,\qquad
	\lambda_{gA}^{ik}=v_g^ia_g^k + a_g^iv_g^k.
	\label{lVA}
}
We use the following Standard Model (SM) parameters: 
$Q_g$ is the electric charge of the particle $g$ in the units of proton charge,
the third component of weak isospin is
$ I_{g}^3=-1/2,\ I_{\nu}^3=+1/2$,
and 
$s_{W}\  (c_{W})$ is the sine (cosine) of Weinberg's angle which is related to the $Z$- and $W$-boson masses 
according to SM prescription as:
\eq{
	c_{W}=\frac{m_{W}}{m_{Z}},\qquad
	s_{W}=\sqrt{1-c_{W}^2}.
	\label{cwsw}
}
Note that in the on-shell renormalization scheme \cite{Bohm:1986rj, Denner:1991kt},
the relations (\ref{cwsw}) are satisfied at every order of perturbation theory.

One can also use symmetrical form of the coupling constants:
\eq{
	g_g^\mp \equiv g_g^{L,R} = v_g^Z \pm a_g^Z,
	\label{Denner:1991kt-var}
}
corresponding to $g_f^\mp$ from Denner's work \cite{Denner:1991kt}.
We omit flavor indexes below since it is not important for the reaction
considered here, i.e. $g_e^\mp=g_\mu^\mp$. 
Thus, in the new notations, all the coupling combinations become symmetric, so we can use the following combinations:
\eq{
	c_0=1+ g^-g^+,\quad c^{\pm}_i=(g^-)^i \pm (g^+)^i \quad \mbox{for}\quad i \geq 1.
}
Let us use the following shortcut for the repeating indexes:
$  b^{ikik}_\mp \equiv  b^{ik}_\mp. $
In Table~\ref{tab.1}, one can find numerical values for $b^{ik}_\mp$ combinations at different
polarizations,
where $L$ and $R$ mean electron polarization degrees $p_{B} = -1$ and $p_{B} = +1$, correspondingly.
A combination $(L+R)/2$ corresponds to unpolarized cross section so we use index
$u$ (unpolarized) for it. This formally occurs at $p_{B} = 0$.
The double indexes of coupling constants appearing in In Table~\ref{tab.1}
(which are needed for box type diagrams) are separated with commas for clearness.
\begin{table}
	\caption{Numerical values of quantities $b^{ik}_\mp$ at different polarizations.}
	\label{tab.1}
	\begin{tabular*}{\columnwidth}{ccccc}
		\hline
		$b^{ik}_\mp$ & $L$ & $R$ & $u=(L+R)/2$ & $L-R$ \\ 
		\hline
		$ b_{\mp}^{\gamma \gamma}$ & ~$ 1 $~       & ~$ 1 $~        &  ~$ 1 $~       & ~$ 0 $   \\        
		$ b_{-}^{\gamma Z}$       & ~$ -0.3568 $~ & ~$ -0.3568 $~  &  ~$ -0.3568 $~ & ~$ 0 $   \\        
		$ b_{+}^{\gamma Z}$       & ~$ +0.4445 $~ & ~$ +0.2864 $~  &  ~$ +0.3654 $~ & ~$ +0.1580 $   \\        
		$ b_{-}^{ZZ}$       & ~$ +0.1273 $~ & ~$ +0.1273 $~  &  ~$ +0.1273 $~ & ~$ 0 $   \\        
		$ b_{+}^{ZZ}$       & ~$ +0.1975 $~ & ~$ +0.0820 $~  &  ~$ +0.1398 $~ & ~$ +0.1155 $   \\        
		$ b_{-}^{ZZ,Z}$       & ~$ -0.0454 $~ & ~$ -0.0454 $~  &  ~$ -0.0454 $~ & ~$ 0 $   \\        
		$ b_{+}^{ZZ,Z}$       & ~$ +0.0878 $~ & ~$ +0.0235 $~  &  ~$ +0.0556 $~ & ~$ +0.0643 $   \\        
		$ b_{-}^{WW,k}$       & ~$ 0 $~ & ~$ 0 $~  &  ~$ 0 $~ & ~$ 0 $   \\        
		$ b_{+}^{WW,\gamma}$  & ~$ +5.0433 $~ & ~$ 0 $~  &  ~$ +2.5216 $~ & ~$ +5.0433 $   \\        
		$ b_{+}^{WW,Z}$       & ~$ +2.2415 $~ & ~$ 0 $~  &  ~$ +1.1208 $~ & ~$ +2.2415 $   \\  
		\hline
	\end{tabular*}
\end{table}
%
These are some useful relations: 
$b^{ik}_\mp = b^{ki}_\mp$, 
$b^{\gamma Z,k}_\mp = b^{Zk}_\mp$
and
$b^{ZZ,\gamma}_\mp = b^{ZZ}_\mp $.
In addition, let us note that
$b^{WW,k}_+|_{L-R} = 2 b^{WW,k}_+|_{u} $.
The latter relation comes from the fact that vector and
axial couplings of fermions with $W$-bosons are the same
(or $g_W^R=0$).

\section{Relative corrections}

\begin{figure}
	\begin{minipage}{\columnwidth}
		\centering
		\subfigure[]{\includegraphics[width=0.48\textwidth]{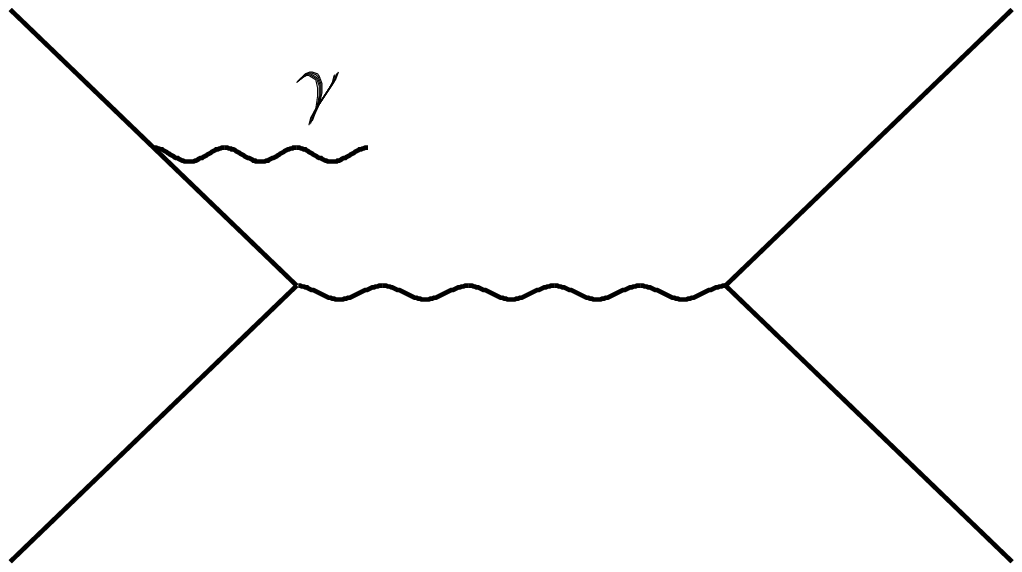}\label{fig.bre1}}
		\hfill
		\subfigure[]{\includegraphics[width=0.48\textwidth]{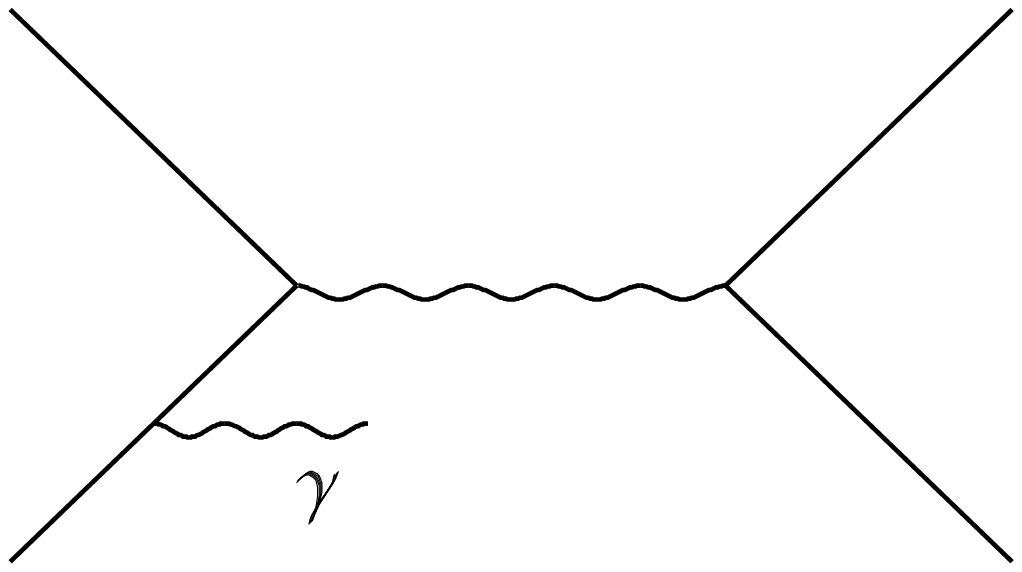}\label{fig.bre2}}
		\\
		\subfigure[]{\includegraphics[width=0.48\textwidth]{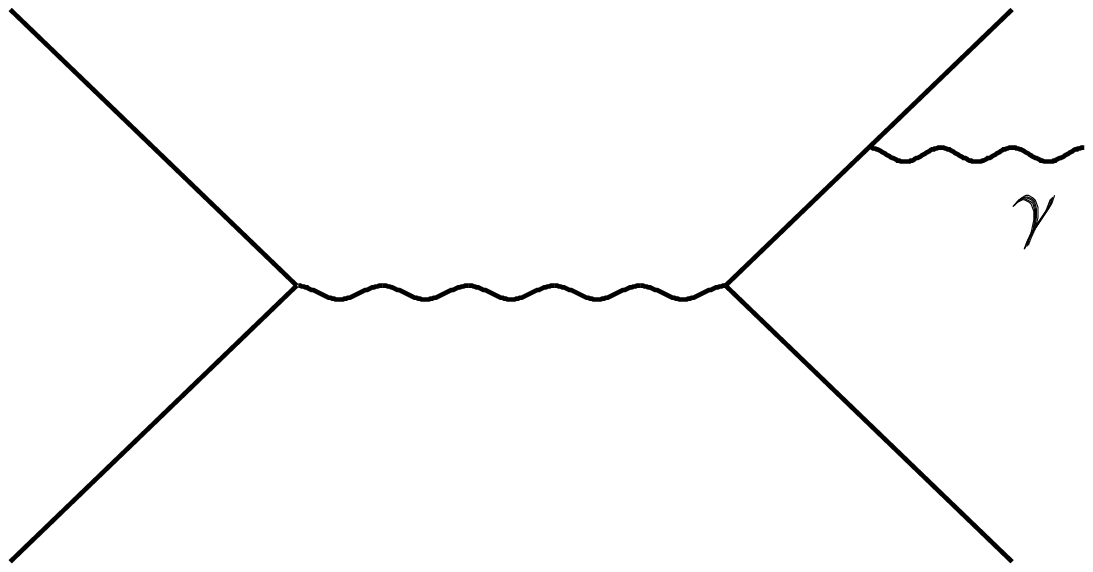}\label{fig.bre3}}
		\hfill
		\subfigure[]{\includegraphics[width=0.48\textwidth]{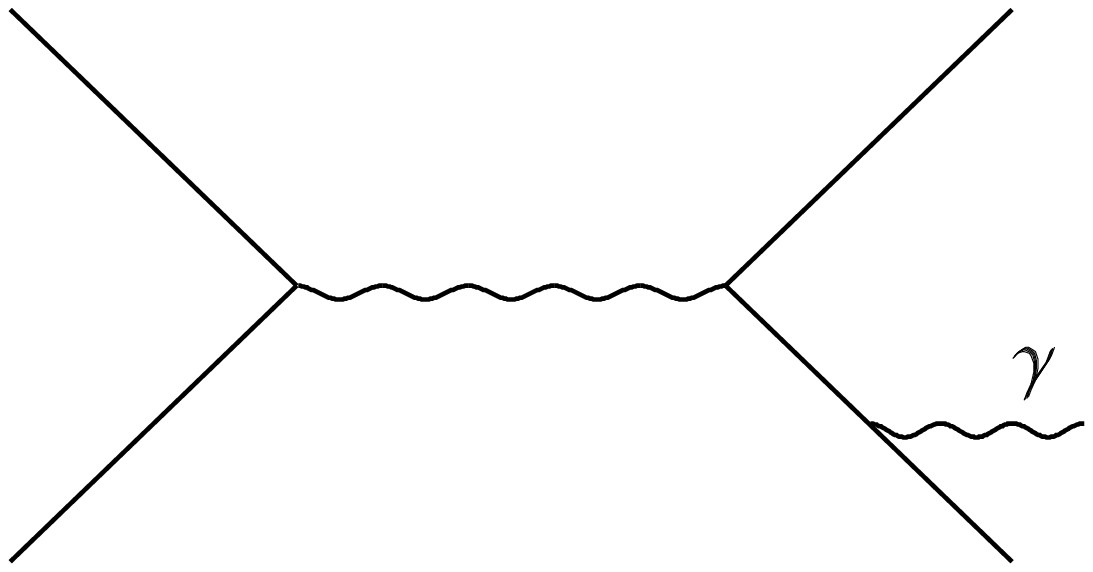}\label{fig.bre4}}
	\end{minipage}
	\caption{The diagrams for photon bremsstrahlung 
		$e^- e^+ \rightarrow \mu^- \mu^+ \gamma$.}
	\label{brem}
\end{figure}

Let us introduce index $C$ to denote the type of the contribution into observables. That is for 
the polarized differential cross sections we can use $\sigma^C_{L,R}$. As for their
combinations
\eq{
	\sigma^C_{u}=\frac{\sigma^C_{L}+\sigma^C_{R}}{2}, 
	\label{s}
}
has the meaning of the unpolarized cross section ($\sigma^C_{u}$).
The polarization asymmetry defined as:
\eq{
	A^C_{LR} =
	\frac{\sigma^C_{L}-\sigma^C_{R}}
	{\sigma^C_{L}+\sigma^C_{R}}.
	\label{A}
}
Index $C$ takes the following values:
$C=\{0, G, \gamma\gamma, \gamma Z, ZZ$, $WW, 1, 0\!+\!1 \}$,
where 
\begin{itemize}
	\item
	0 means Born approximation (Fig.~\ref{fig.Born}),
	\item
	$G$ is the total gauge invariant contribution
	of boson self energies (Fig.~\ref{fig.BSE}),
	vertex functions (Fig.~\ref{fig.V1},~\ref{fig.V2})
	and the part of the cross section corresponding to
	the infrared divergence cancellation including emission of soft photon 
	(see Fig.~\ref{brem}) with the energy below $\omega$,
	that is $G={\rm BSE+Ver+IRD}$,
	\item               
	$\gamma\gamma$, $\gamma Z$, $ZZ$, $WW$ correspond to infrared finite parts of corresponding 
	box type diagrams (Fig.~\ref{fig.BoxDirect},~\ref{fig.BoxCrossed}),
	\item               
	1 stands for full one-loop approximation (NLO),
	\item
	0+1 stands for the calculation within the accuracy of one-loop electroweak
	corrections.
\end{itemize}

The aim of our investigation is the relative corrections of the combination of
differential cross sections which are defined as following:
\eq{
	\delta^C_{\pm} = 
	\frac{\sigma^{C}_{L} \pm \sigma^{C}_{R}}
	{\sigma^{0}_{L} \pm \sigma^{0}_{R}}.
	\label{dpm}
}            
It is clear that these relative corrections are additive, i.e.
\eq{
	\delta^{C_1+C_2}_{\pm}  = \delta^{C_1}_{\pm} + \delta^{C_2}_{\pm},
	\label{additive}
}          
which makes these corrections very convenient for analysis.
With $\delta^C_+$ as the relative correction to unpolarized cross section
and using $\delta^C_-$ one can easily build relative correction to polarized asymmetry:
\eq{
	\delta^C_A=\frac{\delta^C_- - \delta^C_+}{1+\delta^C_+}.
	\label{dAsy}
}            

Let us express equation (\ref{dpm}) in a shorter form,
\eq{
	\delta^C_+ =  \frac{\sigma^{C}_{u}} {\sigma^{0}_{u}},\qquad
	\delta^C_- =  \frac{\sigma^{C}_{L-R}} {\sigma^{0}_{L-R}},
	\label{dpm2}
}            
and calculate the denominators with Born cross section.
In the low energy regime (LE), one has:
\eq{
	\sigma^{0}_{u} &=& \frac{\pi\alpha^2}{s^3}\mu^{\gamma\gamma}|_u= 
	\frac{\pi\alpha^2}{s^3} (t^2+u^2),\\
	\sigma^{0}_{L-R} &=& -\frac{2\pi\alpha^2}{s^2m_Z^2}\mu^{\gamma Z}|_{L-R}= 
	-\frac{2\pi\alpha^2}{s^2m_Z^2} c_2^- \cdot u^2.
	\label{bornLE}
}            
These simple expressions are the result of several approximations:
in the LE-regime, terms from the amplitude with $Z$-boson exchange are suppressed in the unpolarized contribution,
while the contribution from the interference term between $Z$-boson and photon exchange survives in $L-R$ numerator of the asymmetry.
In the high energy (HE) regime, and taking into account that $s \gg m_Z^2$, we get:
\eq{
	\sigma^{0}_{u} &=& \frac{\pi\alpha^2}{s^3} d_0,
	\label{bornHE}\\
	d_0&=&  \sum_{i,j} \mu^{ik}|_u = c_0^2\cdot t^2+(1 + c_2^+ + c_4^+/2)\cdot u^2, \nn\\
	\sigma^{0}_{L-R} &=& \frac{\pi\alpha^2}{s^3} \sum_{i,j} \mu^{ij}|_{L-R}
	= \frac{\pi\alpha^2}{s^3} c_2^- ( 2 + c_2^+ ) \cdot u^2.\nn
}
We note that $c_2^- = {(g_g^-)}^2 - {(g_g^+)}^2 \sim a$, where $a= 1-4s_W^2$ is small.

\section{Infrared divergence cancellation}

Let us first consider the specifically selected part of the one-loop contributions which in sum with soft photon emission
will cancel the infrared divergence. This part is is proportional to Born cross section $\sigma^0$ by definition, with 
procedure outlined in \cite{Aleksejevs:2016tjd}.

So, the cross section with the infrared divergence in soft photon emission contribution
$\frac{\alpha}{\pi} \bigl[ -\delta_1^{\lambda} +R_1 \bigr] \sigma^0 $ will cancel
the infrared-divergent part $ \frac{\alpha}{\pi} \delta_1^{\lambda} \sigma^0$ extracted from $V$-terms (the terms from additional virtual particle contributions), where
\eq{
	\delta_1^{\lambda} &=& 4 B \ln\frac{\lambda}{\sqrt{s}},\qquad \ B= \ln\frac{st}{m_e m_\mu u} - 1,
	\label{dlam}\\
	R_1&=&-4B \ln\frac{\sqrt{s}}{2\omega} 
	- \sum\limits_{g=e,\mu} \Bigl( \ln\frac{m_g^2}{s} +\frac{1}{2}\ln^2\frac{m_g^2}{s} + \frac{\pi^2}{3} \Bigr)
	+\nn\\
	&+&2 \mbox{Li}_2\frac{-t}{u} -2 \mbox{Li}_2\frac{-u}{t}.
}

As for the relative corrections emerging from this part, it is obvious that:
\eq{
	\delta_{\pm}^{\rm IRD} = \frac{\alpha}{\pi} R_1.
}
The cross section with $R_1$ term contains the square of collinear logarithm (CL) which should be absent
in one-loop corrections. Below, we will show that the cancellation of CL square will happen in the sum with vertex-type contributions,
in each relative correction, $\delta_+$ and in $\delta_-$.

\section{Boson self energies}

One of the goals of this paper is to derive the explicit (although approximate) expressions for different contributions to the electroweak corrections.
The gauge-invariant part ($G$-part) is described in in \cite{Aleksejevs:2016tjd}, including the hard photon emission.
The gauge invariance of this part was verified in \cite{Aleksejevs:2010nf}, by demonstrating the same result obtained with different choices of renormalization conditions (by \cite{hollik} and \cite{Denner:1991kt}). All equations  are in the ultra-relativistic approximation and thus
applicable in the region where $\sqrt{s} \gg m_g^2$,  except in a vicinity of resonance where
terms of order $\sim m_g/\Gamma_Z$  would be important.

\begin{figure}
	\begin{minipage}{\columnwidth}
		\centering
		\includegraphics[width=\textwidth]{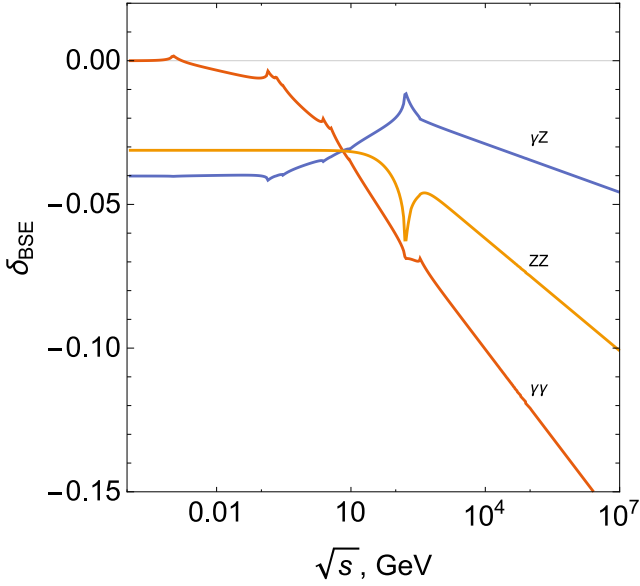}
	\end{minipage}
	\caption{Boson self energies dependence on $\sqrt{s}$.}
	\label{bse}
\end{figure}

We start with the BSE cross section which is infrared-finite:
\eq{
	\sigma^{\rm BSE} 
	&=& \frac{2 \pi \alpha^2}{s} {\rm Re}
	\sum_{i,j,k=\gamma,Z} D_S^{ij}{D^{k}}^* \mu^{ikjk},
	\label{cs-BSE}
}
where
\eq{
	D_S^{ij}=-D^{i} {\hat{\Sigma}}_T^{ij}(s) D^{j},
	\label{D_S}
}
and ${\hat{\Sigma}}_T^{ij}(s)$ is the transverse part of
renormalized self energies of photon, $Z$-boson and $\gamma Z$ mixing.
Fig.~\ref{bse} illustrates the following corrections, corresponding to BSE in Hollik's  renormalization conditions \cite{Bohm:1986rj}:
\eq{
	\delta^{\gamma\gamma}_{\rm BSE} &=&\mbox{Re} {\hat{\Sigma}}_T^{\gamma\gamma}(s) D^{\gamma s},\nn\\
	\delta^{\gamma Z}_{\rm BSE}     &=&\mbox{Re} {\hat{\Sigma}}_T^{\gamma Z}(s) D^{\gamma s},    \nn\\
	\delta^{ZZ}_{\rm BSE}           &=&\mbox{Re} {\hat{\Sigma}}_T^{ZZ}(s) D^{Z s}. \nn
}
For the Belle II kinematics specifically (i.e. at $\sqrt{s}=10.577$ GeV), these corrections are very close to each other:
\eq{
	\delta^{\gamma\gamma}_{\rm BSE} = -0.0361,\ \
	\delta^{\gamma Z}_{\rm BSE}     = -0.0301,\ \
	\delta^{ZZ}_{\rm BSE}           = -0.0317.
}

Let us calculate the relative corrections now.
The cross section in LE-regime has the form:
\eq{
	\sigma^{\rm BSE}_{u, \rm LE} 
	&=& \frac{2 \pi \alpha^2}{s} {\rm Re} D_S^{\gamma\gamma}{D^{\gamma}} \mu^{\gamma\gamma},\nn\\
	\sigma^{\rm BSE}_{L-R, \rm LE} 
	&=& \frac{2 \pi \alpha^2}{s} {\rm Re} \sum_{i,j} D_S^{ij} {D^k}^* \mu^{ikjk}.
	\label{cs-BSE-LE}
}
Substituting these expressions into (\ref{dpm2}), we get:
\eq{
	\delta^{\rm BSE}_{+, \rm LE} &=& -2 \delta^{\gamma\gamma}_{\rm BSE},\\
	\delta^{\rm BSE}_{-, \rm LE} &=& 
	- \delta^{\gamma\gamma}_{\rm BSE} 
	- \frac{2}{c_1^+} \delta^{\gamma Z}_{\rm BSE}
	- \delta^{ZZ}_{\rm BSE}.
	\label{de-BSE-LE}
}
The rest of relative corrections can be obtained in the same way, i.e. by calculating radiative cross section, simplifying and dividing by the Born cross section.
In the HE-regime BSE have a slightly more complicated form, as follows, but still simple enough to be used in quick estimations:
\eq{
	\delta^{\rm BSE}_{+, \rm HE} &=&
	-\frac{1}{d_0} 
	\left( 
	[ 2 c_0 t^2 + (2+c_2^+)u^2] \delta^{\gamma\gamma}_{\rm BSE}
	\right.
	+\nn\\
	&+& 2 c_1^+ [c_0 t^2 + (1- g^+g^- + c_2^+)u^2] \delta^{\gamma Z}_{\rm BSE} + \nn\\
	&+& 
	\left.[2 g^-g^+ c_0 t^2 + (c_2^+ + c_4^+)u^2] \delta^{ZZ}_{\rm BSE}
	\right), \nonumber \\
	\delta^{\rm BSE}_{-, \rm HE} &=& 
	-\frac{2}{c_1^+(2 + c_2^+)} 
	\left(
		c_1^+ \delta^{\gamma\gamma}_{\rm BSE}
		+ 2(c_0^+ + c_2^+) \delta^{\gamma Z}_{\rm BSE}
	+\right.\nn\\
	&+&\left.
	+ (c_0c_1^+ + c_3^+) \delta^{ZZ}_{\rm BSE}
	\right).
	\label{de-BSE-HE}
}

\section{Vertices}

In order to obtain the cross section corresponding to vertices,
\eq{
	\sigma^{\rm Ver} 
	&=& \frac{2\pi \alpha^2}{s} {\rm Re}
	\sum_{i,k=\gamma,Z} D^{i}{D^{k}}^* 
	[ \mu^{F_ikik} + \mu^{ikF_ik} ],
	\label{cs-Ver}
}
we follow \cite{Bohm:1986rj} and use the renormalized form factors
to replace coupling constants. The form factors decomposed into two terms:
\eq{
	v_g^{\gamma (Z)} \rightarrow v_g^{F_{\gamma (Z)}},\qquad
	a_g^{\gamma (Z)} \rightarrow a_g^{F_{\gamma (Z)}},\qquad
	C_g =\frac{\alpha }{4\pi}  \Lambda_{1,g}^{\gamma}
}
where for a photon one has:
\eq{
	v_g^{F_\gamma} &=&
	C_g v_g^{\gamma}
	+
	\frac{\alpha }{4\pi}   \Bigl[ 
	\bigl({(v_g^Z)}^2 + {(a_g^Z)}^2 \bigr) \Lambda_2^Z
	+ \frac{3}{4s_{W}^2} \Lambda_3^W \Bigr],
	\label{HV1}\\
	a_g^{F_\gamma} &=&
	C_g  a_g^{\gamma}
	+
	\frac{\alpha }{4\pi}  \Bigl[ 
	2v_g^Za_g^Z \Lambda_2^Z
	+ \frac{3}{4s_{W}^2} \Lambda_3^W \Bigr],
	\label{HV2}
}
while for $Z$-boson:
\eq{
	v_g^{F_Z}  &=& 
	C_g  v_g^Z
	+
	\frac{\alpha }{4\pi}   \left[ 
	v_g^Z \bigl({(v_g^Z)}^2 + 3{(a_g^Z)}^2 \bigr) \Lambda_2^Z +
	\right.\nn\\
	&+&\left.
	\frac{1}{8s_{W}^3c_{W}} \Lambda_2^W  -
	\frac{3c_{W}}{4s_{W}^3} \Lambda_3^W \right],
	\label{HV3}
	\\
	a_g^{F_Z} &=& 
	C_g  a_g^Z
	+
	\frac{\alpha }{4\pi}  \left[ 
	a_g^Z \bigl( 3{(v_g^Z)}^2 + {(a_g^Z)}^2 \bigl) \Lambda_2^Z +
	\right.\nn\\
	&+&\left.
	\frac{1}{8s_{W}^3c_{W}} \Lambda_2^W  -
	\frac{3c_{W}}{4s_{W}^3} \Lambda_3^W \right].
	\label{HV4}
}

The function $\Lambda_{1,g}^\gamma$ which enters into factor $C_g$ describes the contribution of
the triangle diagram with a photon exchange, $\Lambda_2$ is for diagrams with a massive boson -- $Z$ or $W$, 
and $\Lambda_3$ is for diagrams with three-boson vertex -- $WW\gamma$ or $WWZ$.
These complex functions can be found in \cite{hollik}.
A real part on the first function for the $s$-channel contains collinear logarithms:
\eq{
	{\rm Re} \Lambda_{1,g}^{\gamma} &=& -2 \ln\frac{s}{\lambda^2} \Bigl( \ln\frac{s}{m_g^2}-1 \Bigr) 
	+ \ln\frac{s}{m_g^2} + 
	\nn\\
	&+& \ln^2\frac{s}{m_g^2} +  4 \Bigl(\frac{\pi^2}{3}-1 \Bigr).
	\label{Lam1}
}          
In the LE-regime, we have:
\eq{
	\Lambda_2^{B} = \Bigl( \frac{2}{3} \ln\frac{m_{B}^2}{s} + \frac{11}{9} \Bigr) \frac{s}{m_{B}^2},\qquad
	\Lambda_3^{W} = -\frac{5}{27} \frac{s}{m_{W}^2},
	\label{Lam23LE}
}      
and at the high energies it has the form:
\eq{
	\Lambda_2^{B} &=& -\ln^2\frac{m_{B}^2}{s} -3 \ln\frac{m_{B}^2}{s} + \frac{\pi^2}{3} - \frac{7}{2},
	\nn\\
	\Lambda_3^{W} &=&  -\frac{1}{3}\ln\frac{m_{W}^2}{s} + \frac{5}{6}.
	\label{Lam23HE}
}

Let us present relative infrared-finite corrections from the vertex diagrams.
For that, as it was shown in \cite{Aleksejevs:2016tjd}), we make the following replacement in the form factors: $\lambda \rightarrow \sqrt{s}$.
For the LE-regime, one has:
\eq{
	\delta^{\rm Ver}_{+, \rm LE} &=&
	\frac{\alpha}{2\pi} 
	\Bigl(
	(\Lambda_{1,e}^{\gamma}+\Lambda_{1,\mu}^{\gamma})|_{\lambda \rightarrow \sqrt{s}}
	+ c_2^+\Lambda_2^{Z}+ \frac{3}{2s_W^2} \Lambda_3^{W} \Bigr),\\
	\delta^{\rm Ver}_{-, \rm LE} &=&
	\frac{\alpha}{2\pi} 
	\Bigl(
	(\Lambda_{1,e}^{\gamma}+\Lambda_{1,\mu}^{\gamma})|_{\lambda \rightarrow \sqrt{s}}
	- \Delta_1 \Bigr).
	\label{de-Ver-LE}
}
where a dominant contribution is coming from the photon 
vertices with an additional heavy vector boson, and has the following form:
\eq{
	\Delta_1 = \frac{m_Z^2}{s} \Bigl( \Lambda_2^{Z} + \frac{3}{2 c_2^- s_W^2} \Lambda_3^{W} \Bigr) 
	= \frac{2}{3} \ln\frac{m_{Z}^2}{s} + \frac{11}{9} - \frac{10}{9a}.
}
This contribution is important, since it contains logarithm which increases with decreasing $s$, and because it has a big value of $1/a$.
In the HE-regime, we obtain:  
\eq{
	\delta^{\rm Ver}_{+, \rm HE} &=&
	\frac{\alpha}{2\pi} (\Lambda_{1,e}^{\gamma}+\Lambda_{1,\mu}^{\gamma})|_{\lambda \rightarrow \sqrt{s}}
	+\nn\\
	&+& \frac{\alpha}{16 \pi s_W^2 c_W^2 d_0}
	\Bigl(
	\Bigl[ \frac{1}{c_W^2} \Lambda_2^{W} +\frac{8s_W^4-4s_W^2+1}{2c_W^4} \Lambda_2^{Z} \Bigr] t^2 + \nonumber\\
	&+& \left[ \frac{-1+2s_W^2}{4s_W^4c_W^2} \Lambda_2^{W} 
	+ \frac{3}{2s_W^4} \Lambda_3^{W}
	\right.+\nn\\
	&+&\left. \frac{64s_W^8+4s_W^4-4s_W^2+1}{8s_W^4c_W^4} \Lambda_2^{Z} \right] u^2
	\Bigr), \nonumber \\
	\delta^{\rm Ver}_{-, \rm HE} &=&
	\frac{\alpha}{2\pi} (\Lambda_{1,e}^{\gamma}+\Lambda_{1,\mu}^{\gamma})|_{\lambda \rightarrow \sqrt{s}} 
	+ \frac{\alpha}{2 \pi s_W^2 a (1+4s_W^2)} \times
	\nn\\
	&\times& \left(
	(2s_W^2-1) \Lambda_2^{W} + 6 c_W^2 \Lambda_3^{W}
	\right.+\nn\\
	&+&\left.
	+ \frac{-64s_W^8+4s_W^4-4s_W^2+1}{2c_W^2} \Lambda_2^{Z} 
	\right).
	\label{de-Ver-HE}
}
Finally, summing up the infrared-divergent and boson self-energy parts of corrections, we can
demonstrate that the square of collinear logarithm cancels in the final result.

\begin{figure}
	\begin{minipage}{\columnwidth}
	\centering
	\includegraphics[width=\textwidth]{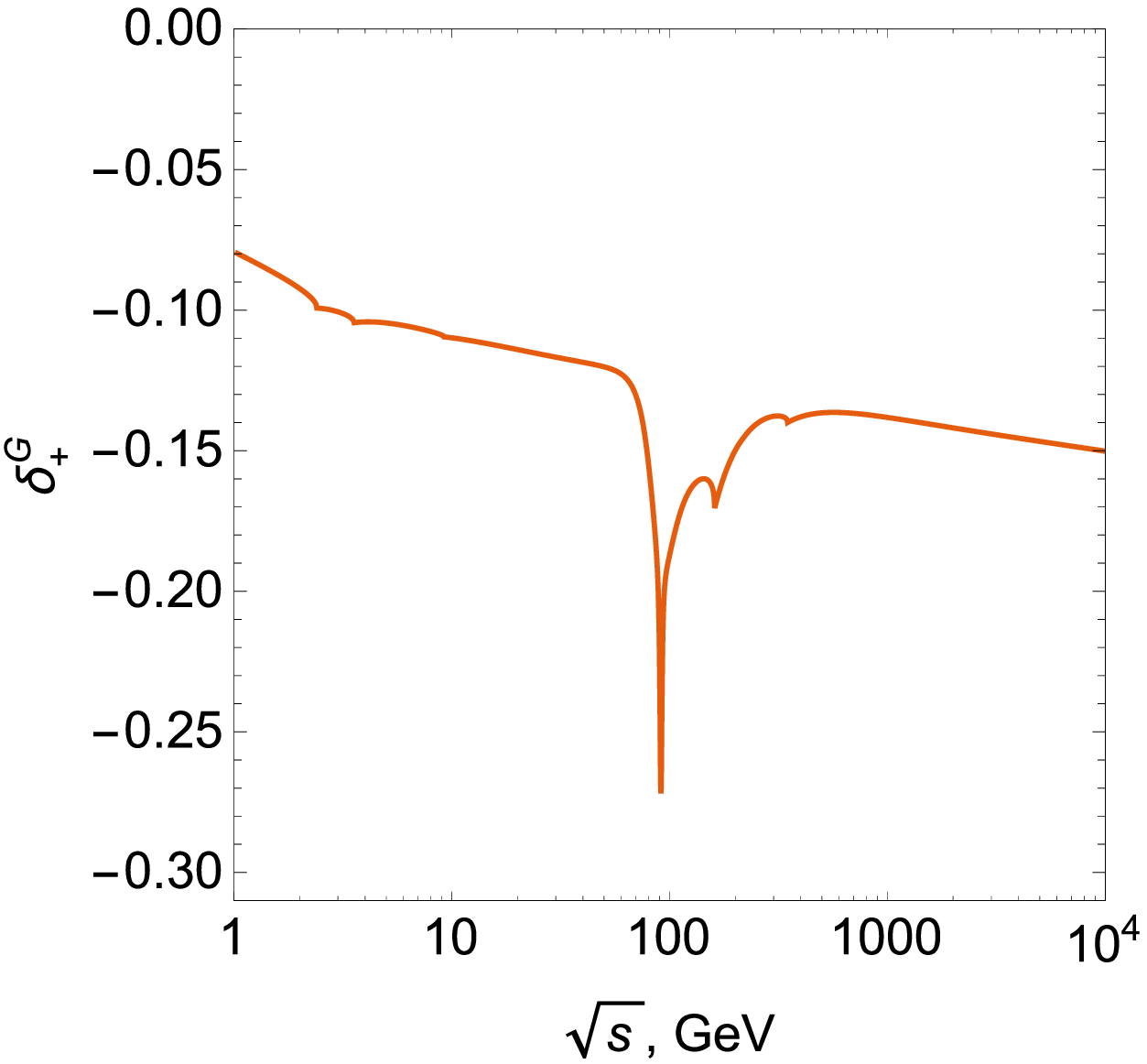}
	\\
	\includegraphics[width=\textwidth]{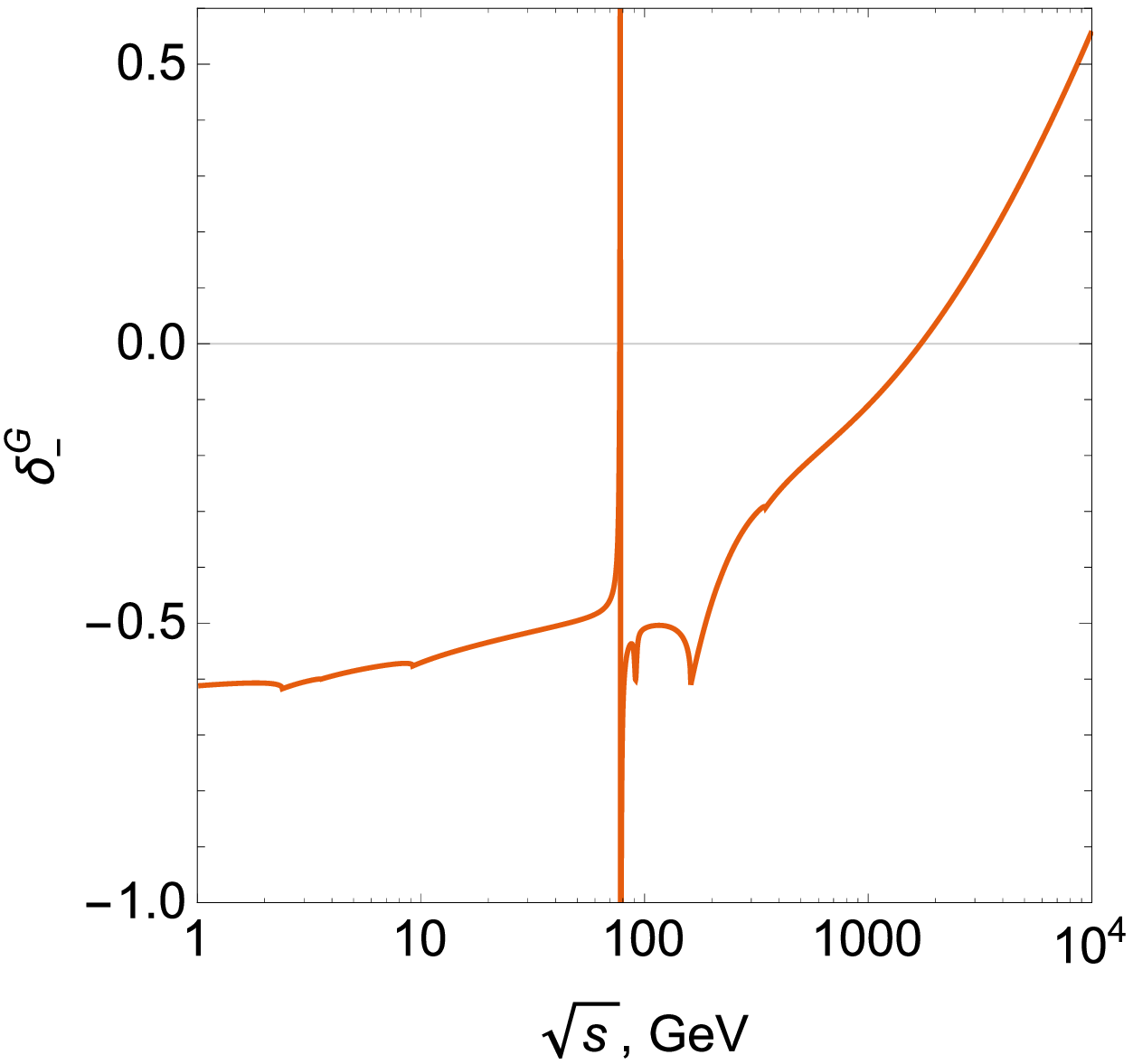}
	\end{minipage}
	\caption{Relative corrections induced by gauge invariant set (BSE+Ver+IRD) at $\theta=90^\circ$.}
	\label{g}
\end{figure}

 Figure~\ref{g} shows numerical results for
gauge invariant set (BSE+Ver+IRD) at $\theta=90^\circ$.

\section{Box diagrams}

Here we provide detailed results for box-type contributions.
Note that
$\gamma\gamma$, $\gamma Z$ and $ZZ$-boxes contain both direct and crossed legs parts,
while $WW$-box has only direct diagram.
The latter feature comes from the electric charge conservation law, which in case lets say
$u\bar u \rightarrow \mu^-\mu^+$ scattering (see, for instance, \cite{Zykunov:2005tc}) changes the effect, and $WW$-box diagram in this case has only crossed-legs term.
The general rule for getting a crossed box from a known direct box is well-known, and in our notations has the form:
\eq{
	\sigma^{C}_{\rm box} = - \sigma^{D}_{\rm box}|_{t \leftrightarrow u,\ b_+ \leftrightarrow b_- }.
	\label{DC}
}
From now on, we will be taking only real part of 
the interference term in our expressions for cross sections.

The infrared-finite part of $\gamma\gamma$-box in $s$-channel can be found in \cite{Zykunov:2006yb,Zykunov:2005tc,Aleksejevs:2016tjd}. From \cite{Aleksejevs:2016tjd} (with misprints corrected):
\eq{
	\sigma^D_{\gamma\gamma} 
	&=& -\frac{\alpha^3}{s} 
	\sum_{k=\gamma,Z} {D^{k}}^* f^{\gamma,k},
	\label{cs-gg}
}
\eq{
	f^{i, k}
	= b_-^{i, k} \Bigl( \frac{t^2+u^2}{2s} \ln^2\frac{s}{|u|} + t \ln\frac{s}{|u|} \Bigr) 
	+ b_+^{i, k} \frac{u^2}{s} \ln^2\frac{s}{|u|}.
}
That works for arbitrary energies, and gives the following expression
for relative corrections:
\eq{
	\delta^{\gamma\gamma-{\rm box}}_{+} &=&
	-\frac{\alpha}{2\pi} \frac{1}{t^2+u^2} 
	\left( 
	(t^2+3u^2) \ln^2\frac{s}{u}+ 2st \ln\frac{s}{u} -
	\right.\nn\\
	&-& \left. (t \leftrightarrow u)
	\frac{}{}\right), \nn \\
	\delta^{\gamma\gamma-{\rm box}}_{-} &=&
	-\frac{\alpha}{2\pi} 
	\Bigl( 
	\ln\frac{s}{u} - \frac{t^2+u^2}{2u^2}\ln^2\frac{s}{t}-\frac{s}{u}\ln\frac{s}{t}
	\Bigr).
	\label{de-gg-box}
}

The approach for obtaining expressions for the amplitudes 
with at least one massive boson at the energies below $Z$-resonance (in LE-regime)
is explained in \cite{Aleksejevs:2010ub}.
By applying this method in \cite{Aleksejevs:2016tjd} for a direct box diagram, we got
expressions for the low (below $Z$-resonance) energies (LE-regime).
Let us present here their infrared finite parts using notations of this paper:
%
\eq{
	\sigma^D_{\gamma Z} 
	&=& -\frac{\alpha^3}{s} \Bigl( \ln\frac{-t}{m_Z^2} - 1 \Bigr) D^{Z}
	\times\nn\\
	&\times&
	\sum_{k=\gamma,Z} {D^{k}}^* 
	\bigl( 
	b_-^{Z,k} 4t^2 + b_+^{Z,k} u^2
	\bigr),
	\label{cs-gZ}
	\\
	\sigma^D_{ZZ} 
	&=& -\frac{\alpha^3}{2s m_Z^2} 
	\sum_{k=\gamma,Z} {D^{k}}^* 
	\bigl( 
	b_-^{ZZ,k} 4t^2 + b_+^{ZZ,k} u^2
	\bigr).
	\label{cs-ZZ}
}
Finally, we can write out the coupling constants for $ZZ$- and $WW$-boxes:
\eq{
	v^{ZZ}&=&{(v^Z)}^2+{(a^Z)}^2,\qquad a^{ZZ}=2v^Za^Z,\nn\\
	v^{WW}&=&a^{WW}=1/4s_W^2.
}

To calculate the box diagrams in the HE-regime, we
use the asymptotic method of \cite{Zykunov:2005tc}. Then, for the direct box (infrared-finite parts only), we get:
\eq{
	\sigma^D_{\gamma Z} 
	&=& - 2 \frac{\alpha^3}{s} 
	\sum_{k=\gamma,Z} {D^{k}}^* 
	\Bigl( 
	D^\gamma  \mu^{Z,k} L_2(-m_Z^2/t)
	+ f^{Z,k}
	\Bigr),
	\label{HE-cs-gZ}
	\\
	\sigma^D_{ZZ} 
	&=& -\frac{\alpha^3}{s} 
	\sum_{k=\gamma,Z} {D^{k}}^* 
	\Bigl( 
	2 D^Z  \mu^{ZZ,k}  L_2(-m_Z^2/t)
	+ f^{ZZ,k}
	\Bigr),
	\label{HE-cs-ZZ}
}
where
\eq{
L_2(\epsilon) = \int_0^1 dx \frac{1}{1-x-\epsilon} \ln\frac{1-x}{\epsilon}
\approx  \frac{\pi^2}{3} + \frac{1}{2} \ln^2\frac{1}{\epsilon}.
}

To obtain the direct $WW$-box,we only need to do the obvious replacements
in (\ref{cs-ZZ}) and (\ref{HE-cs-ZZ}):
$Z \rightarrow W$, $ZZ \rightarrow WW$.
Now we can calculate the relative corrections
for box diagrams with one and two massive bosons.

In the LE-regime, the relative corrections are,\\
for $\gamma Z$-box:
\eq{
	\delta^{\gamma Z-{\rm box}}_{+} &=&
	\frac{\alpha}{8\pi s_W^2c_W^2} \frac{s}{m_Z^2} \frac{1}{t^2+u^2} 
	\times\nn\\
	&\times&
	\Bigl( 
	4s_W^2(2s_W^2-1)  (4\ln\frac{-t}{m_Z^2}-\ln\frac{-u}{m_Z^2}-3 ) t^2 \nonumber\\
	&+&  (8s_W^4-4s_W^2+1) (\ln\frac{-t}{m_Z^2}-4\ln\frac{-u}{m_Z^2}+3) u^2
	\Bigr), \nonumber \\
	\delta^{\gamma Z-{\rm box}}_{-} &=&
	\frac{\alpha}{2\pi} 
	\Bigl( 
	4\ln\frac{-u}{m_Z^2} - \ln\frac{-t}{m_Z^2}-3
	\Bigr).
	\label{de-gZ-box-LE}
}
for $ZZ$-box:
\eq{
	\delta^{ZZ-{\rm box}}_{+} &=&
	\frac{3\alpha}{4\pi} \frac{s}{m_Z^2} \frac{1}{t^2+u^2} 
	\Bigl( 
	-2 (g^-g^+)^2 t^2  + c_4^+ u^2 
	\Bigr), \nonumber \\
	\delta^{ZZ-{\rm box}}_{-} &=&
	-\frac{3\alpha}{4\pi} c_2^+.
	\label{de-ZZ-box-LE}
}
and for $WW$-box 
\eq{
	\delta^{WW-{\rm box}}_{+} &=&
	-\frac{\alpha}{16 \pi s_W^4} \frac{s}{m_W^2} \frac{u^2}{t^2+u^2},
	\nonumber \\
	\delta^{WW-{\rm box}}_{-} &=&
	\frac{\alpha}{16 \pi s_W^4 c_W^2 c_2^-}.
	\label{de-WW-box-LE}
}

In the HE-regime, the relative corrections
$$ L_1 =  \ln^2\frac{-t}{m_Z^2} - \ln^2\frac{-u}{m_Z^2},
$$
will have a typical structure showing that collinear logarithm power is reduced to one, so,\\
for $\gamma Z$-box:
\eq{
	\delta^{\gamma Z-{\rm box}}_{+} &=&
	-\frac{\alpha}{2\pi} \frac{1}{d_0} 
	L_1
	\Bigl(
	2g^-g^+c_0 t^2 + (c_2^+ + c_4^+) u^2
	\Bigr), \nonumber \\
	\delta^{\gamma Z-{\rm box}}_{-} &=&
	-\frac{\alpha}{\pi} 
	L_1 \frac{1+c_2^+}{2+c_2^+}.
	\label{de-gZ-box-HE}
}
for $ZZ$-box:
\eq{
	\delta^{ZZ-{\rm box}}_{+} &=&
	-\frac{\alpha}{2\pi} \frac{1}{d_0} L_1
	\Bigl( 
	2 (g^-g^+)^2 c_0 t^2  + (c_4^+ + c_6^+) u^2 
	\Bigr), \nonumber \\
	\delta^{ZZ-{\rm box}}_{-} &=&
	-\frac{\alpha}{\pi} L_1 \frac{ (g^-g^+)^2 + c_2^+ + c_4^+ }{2+c_2^+}.
	\label{de-ZZ-box-HE}
}
and for $WW$-box 
\eq{
	\delta^{WW-{\rm box}}_{+} &=&
	-\frac{\alpha}{4 \pi } \frac{1+{(g^-)}^2}{s_W^4} \frac{u^2}{d_0}  
	L_2(-m_W^2/t),
	\nonumber \\
	\delta^{WW-{\rm box}}_{-} &=&
	-\frac{\alpha}{2 \pi}  \frac{1+{(g^-)}^2}{s_W^4 c_2^- (2+ c_2^+)}
	L_2(-m_W^2/t).
	\label{de-WW-box-HE}
}

\begin{figure}
	\begin{minipage}{\columnwidth}
		\centering
		\includegraphics[width=0.49\textwidth]{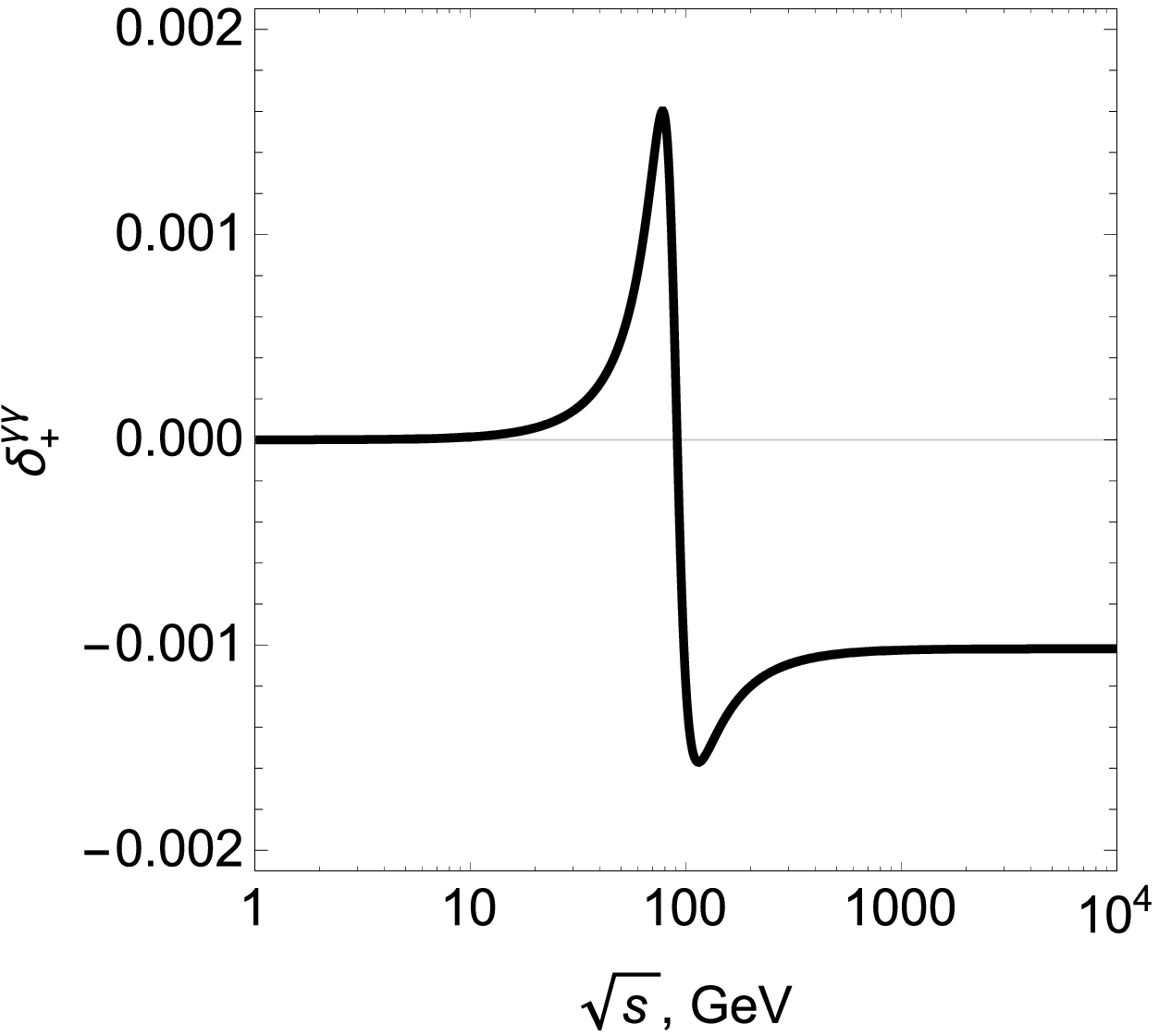}
		\hfill
		\includegraphics[width=0.49\textwidth]{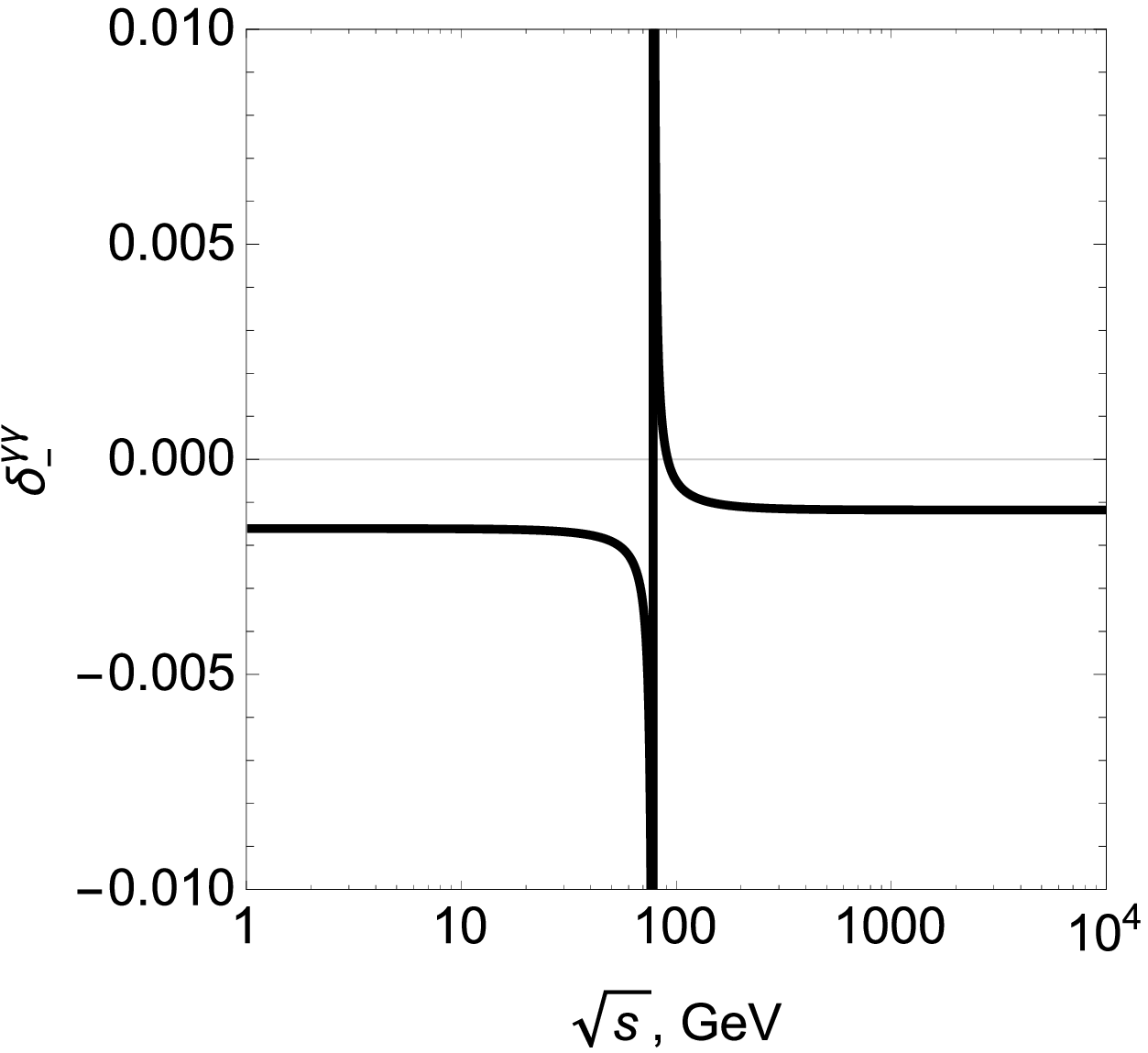}
		\\
		\includegraphics[width=0.49\textwidth]{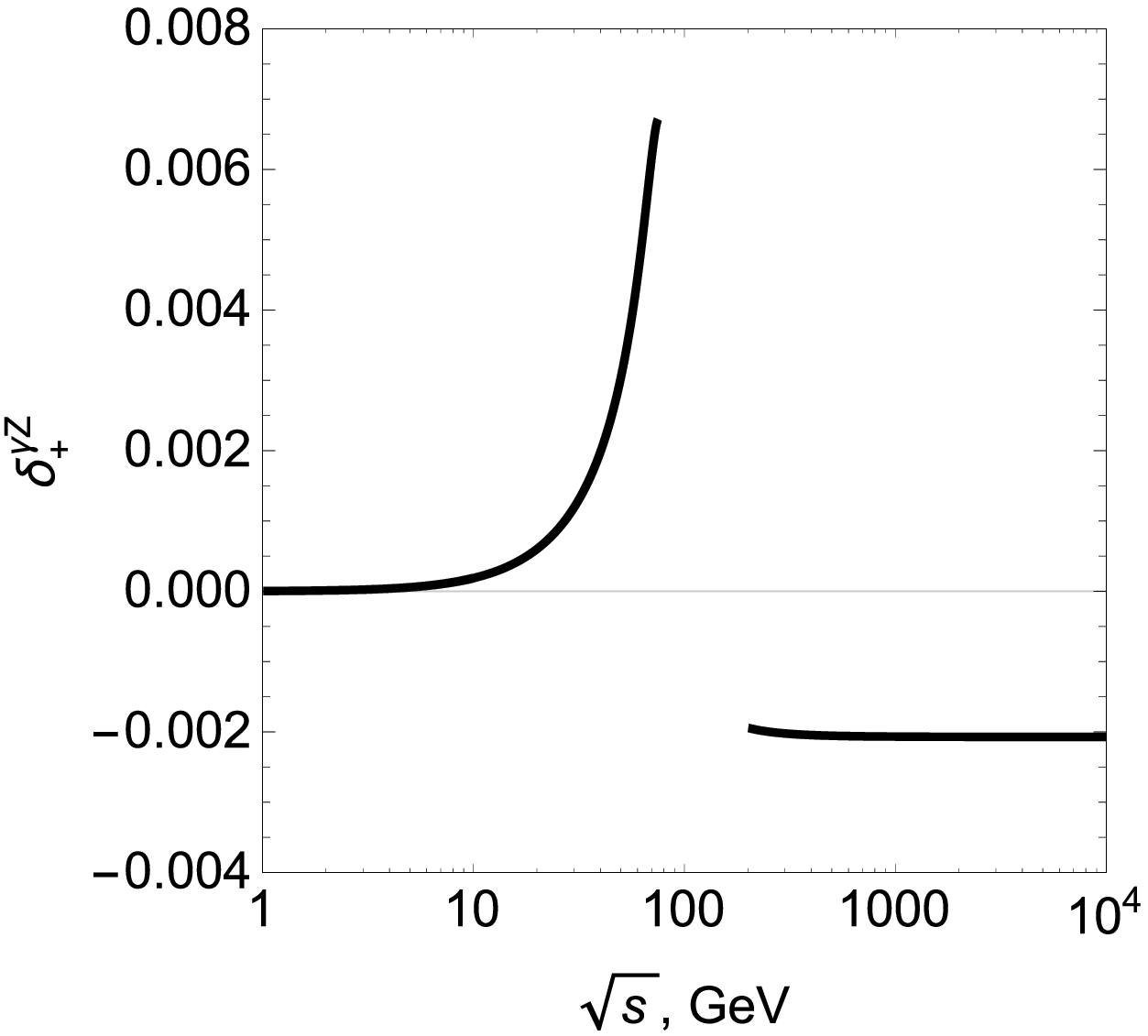}
		\hfill
		\includegraphics[width=0.49\textwidth]{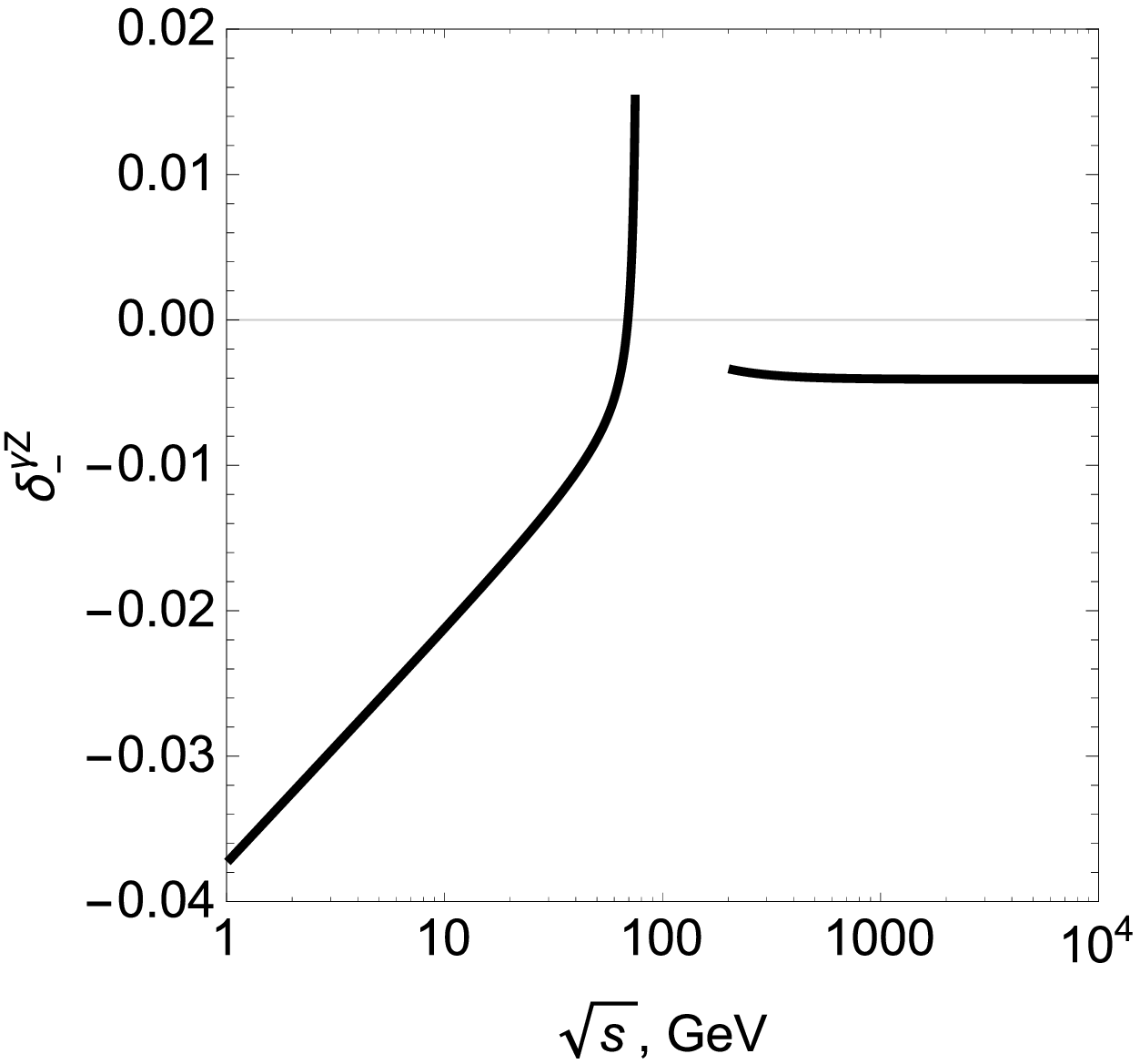}
		\\
		\includegraphics[width=0.49\textwidth]{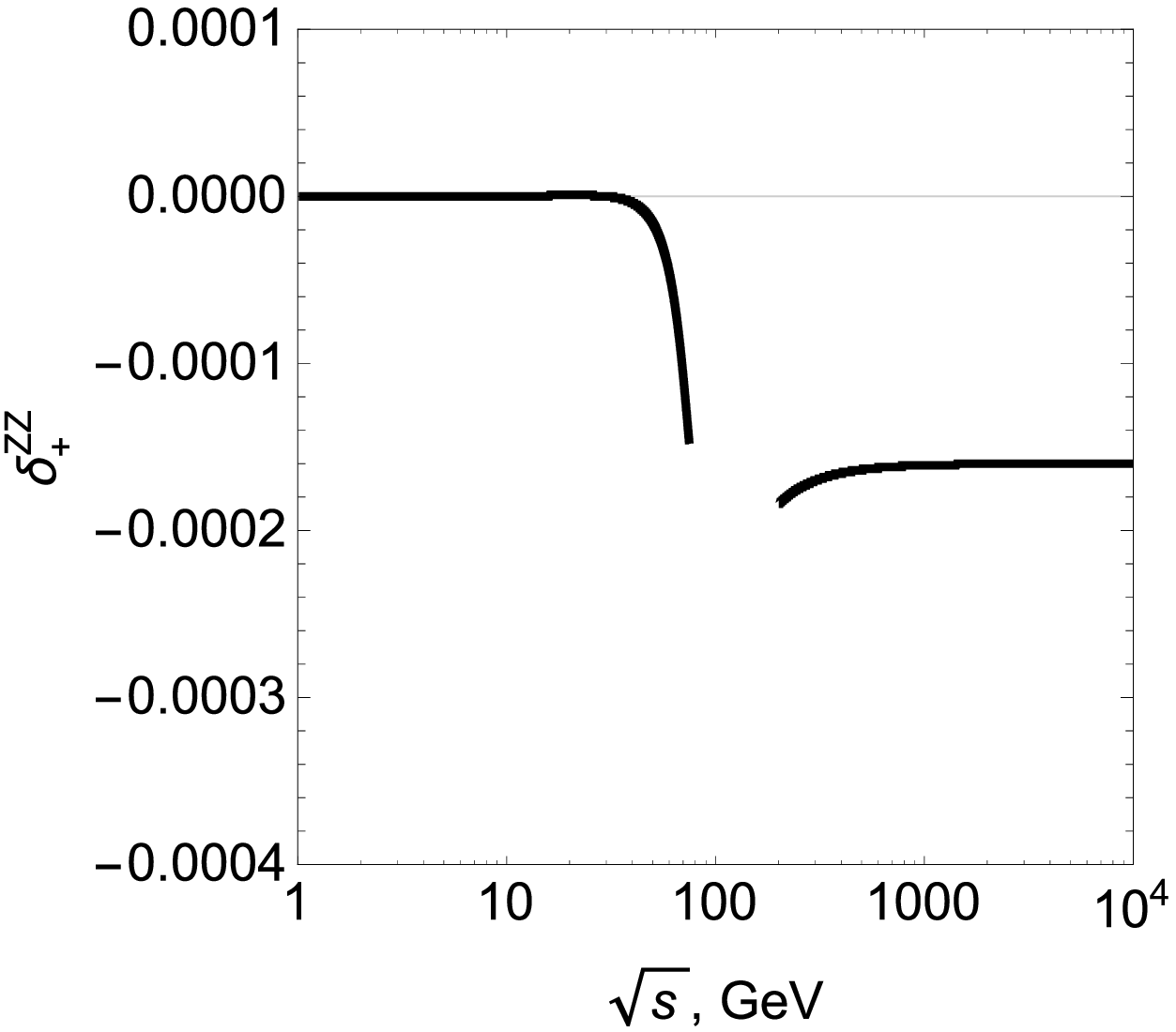}
		\hfill
		\includegraphics[width=0.49\textwidth]{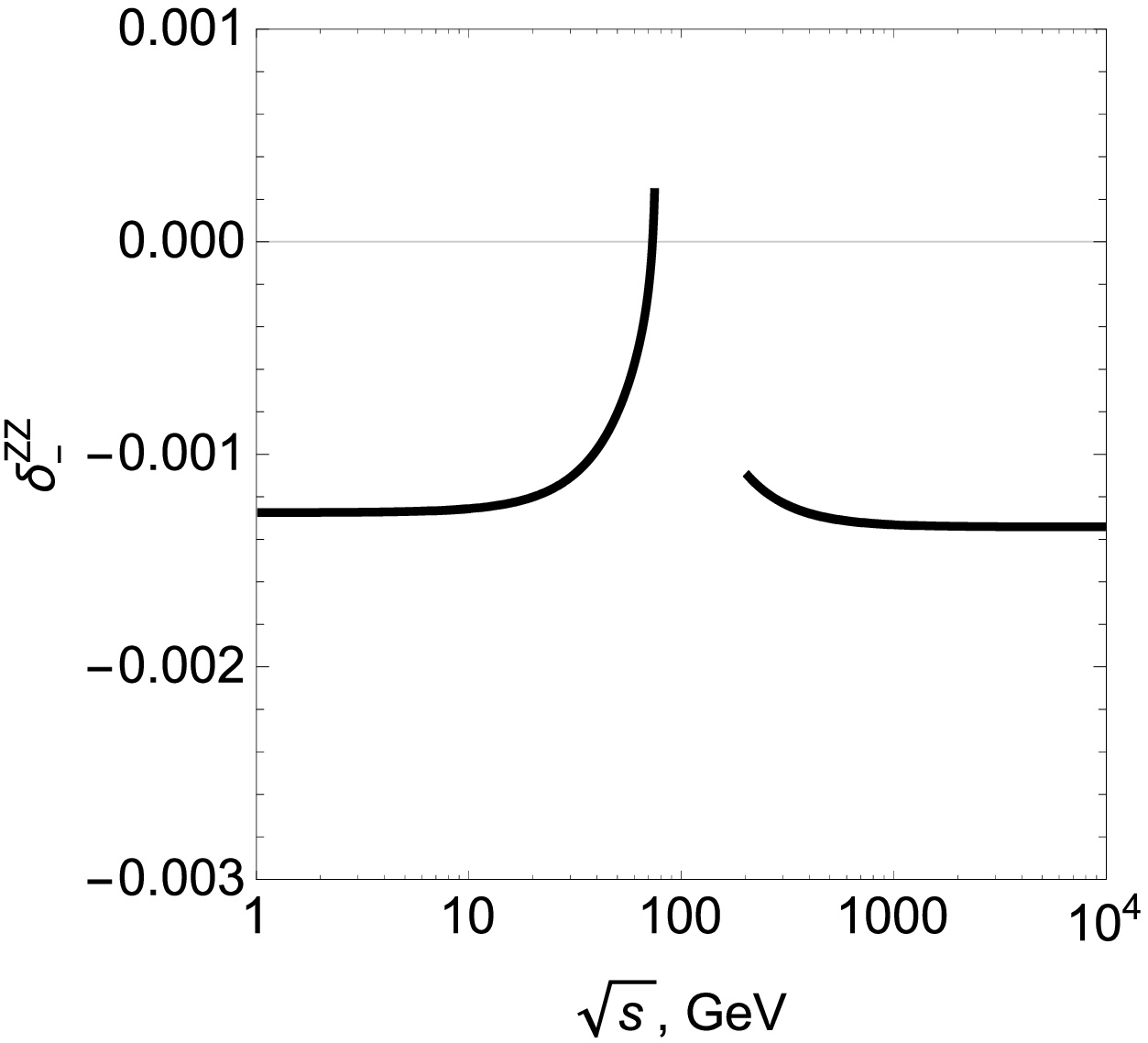}
	\end{minipage}
	\caption{Relative corrections induced by $\gamma\gamma$, $\gamma Z$ and $ZZ$-boxes at $\theta=90^\circ$.}
	\label{gZ}
\end{figure}

\begin{figure}
	\begin{minipage}{\columnwidth}
		\centering
		\includegraphics[width=0.49\textwidth]{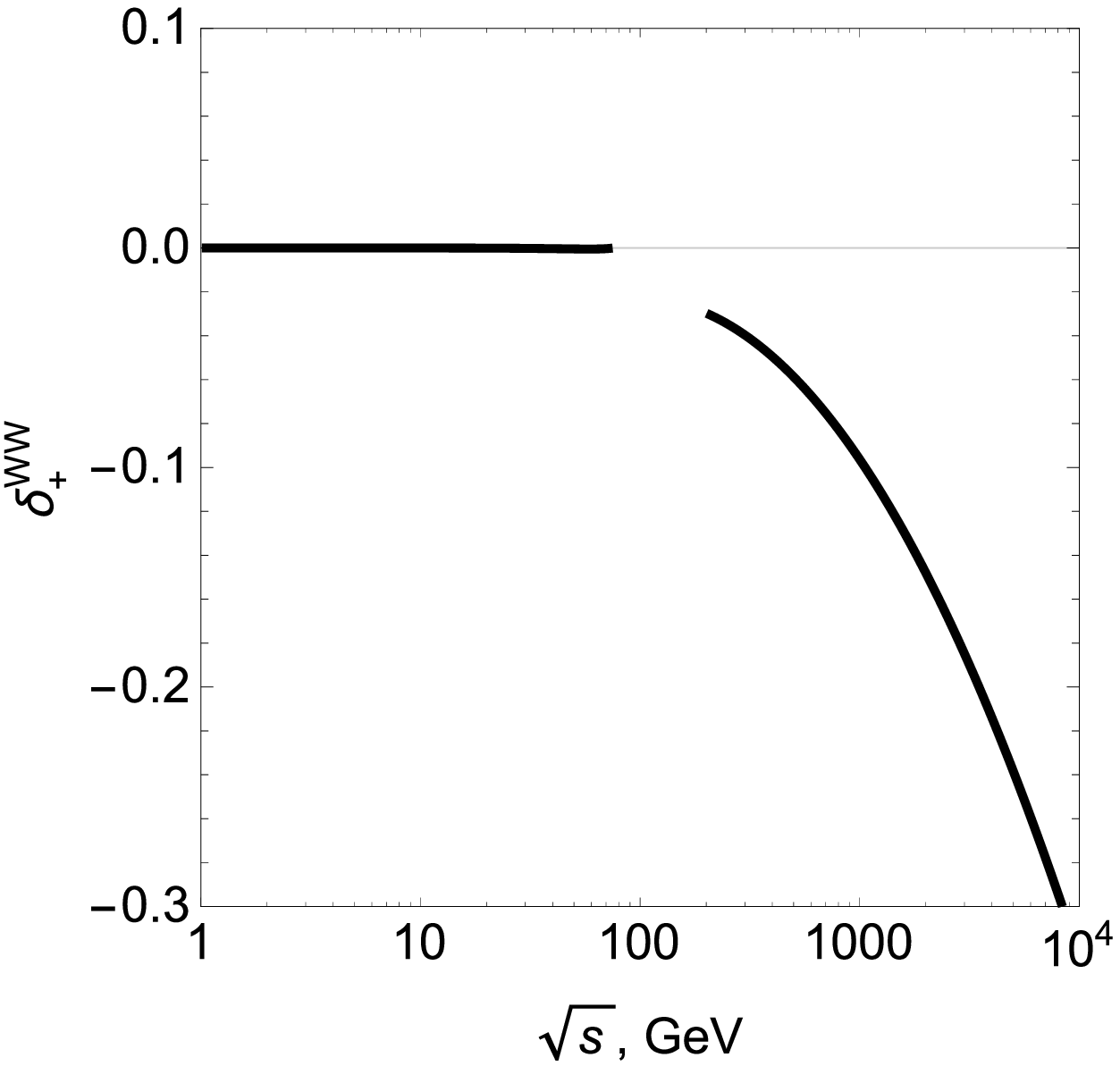}
		\hfill
		\includegraphics[width=0.49\textwidth]{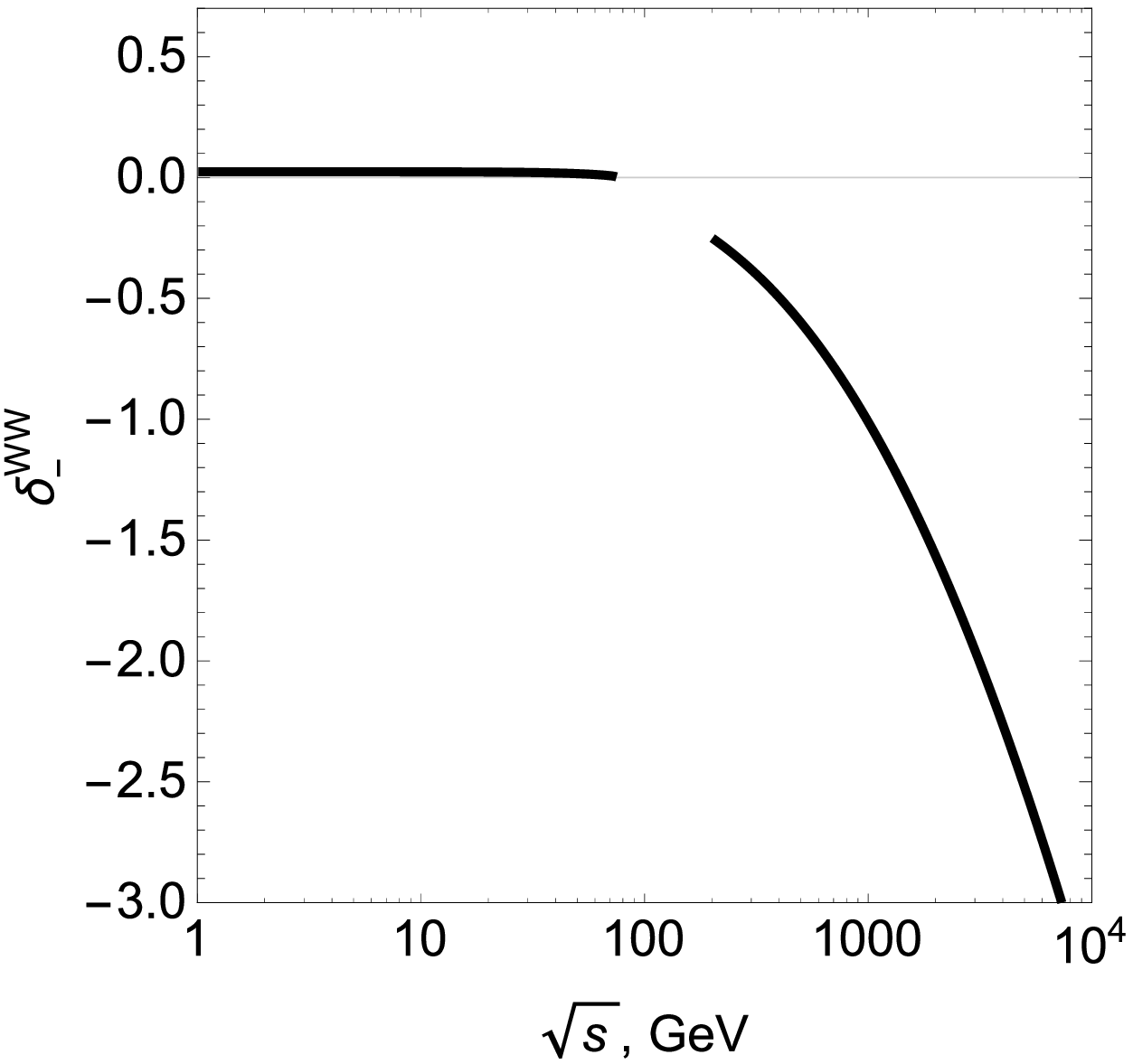}
	\end{minipage}
	\caption{Relative corrections induced by $WW$-box at $\theta=90^\circ$.}
	\label{WW}
\end{figure}

\begin{figure}
	\begin{minipage}{\columnwidth}
		\centering
		\includegraphics[width=0.49\textwidth]{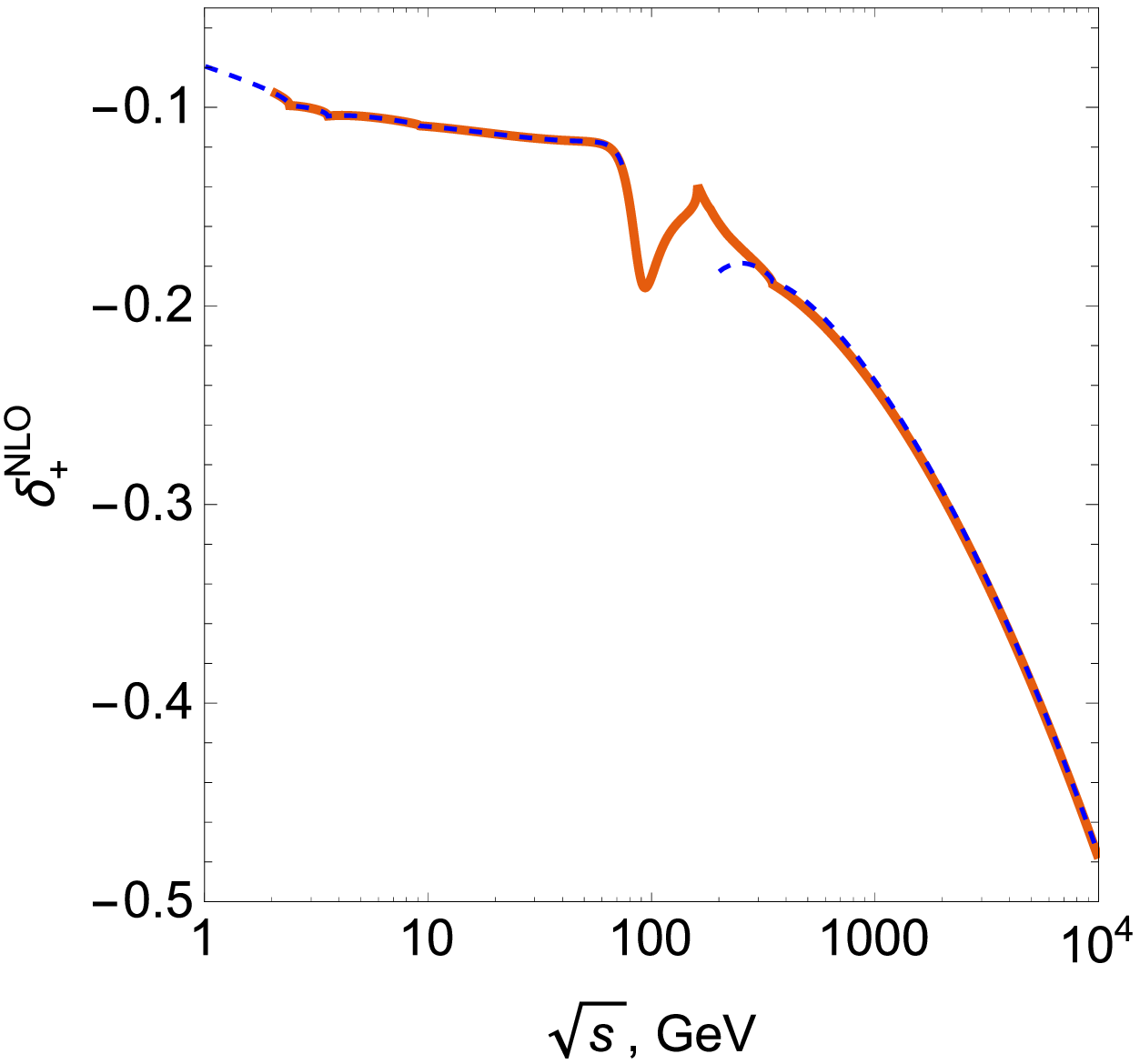}
		\hfill
		\includegraphics[width=0.49\textwidth]{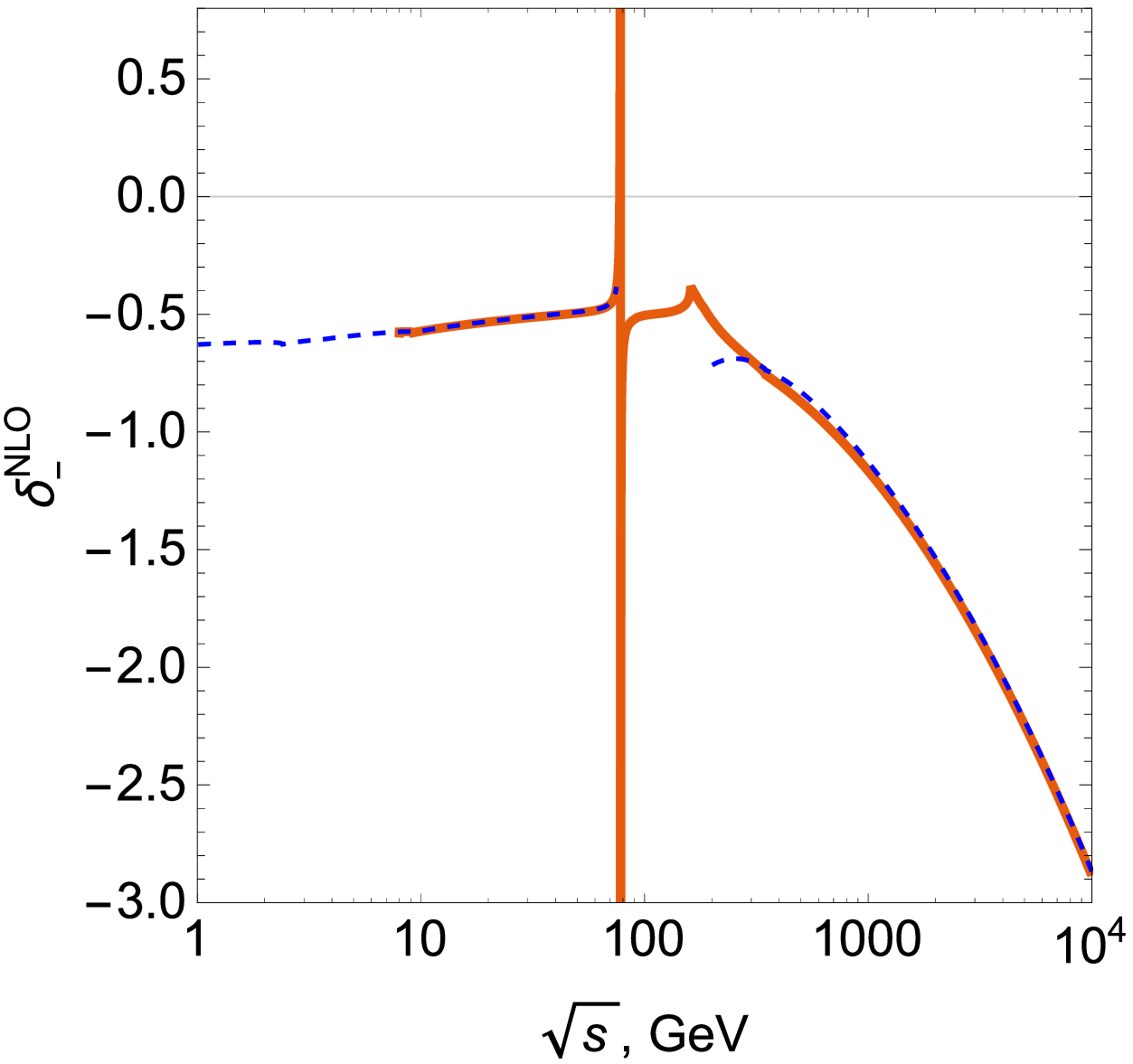}
	\end{minipage}
	\caption{Total relative NLO corrections at $\theta=90^\circ$.}
	\label{NLO}
\end{figure}

\section{Analysis and Conclusions}

We evaluate a complete set of electroweak radiative correction to the parity-violating asymmetry 
in $e^- e^+ \rightarrow mu^- mu^+ (\gamma)$ at one loop, i.e. the next-to-the-leading order (NLO) level and demonstrate that they are fully under control. 
Our first approach, more time-honored and better-tested, relies on calculations "on paper" with reasonable approximations well-supported in the literature, while our second approach, more novel, relies on program packages FeynArts, FormCalc, LoopTools and Form. We demonstrate that in the high- and low-energy regions, well below and above Z-resonance, correspondingly, our numerical results obtained with these two independent approaches are in a very good agreement. 

The goal of this work is to provide the experimental community with options to better suit their needs, depending on the timeliness and the required precision. Clearly, for a full data analysis of ultra-precision experiments such as MOLLER, P2 and Belle-II, it would be essential to retain the maximum precision by evaluating a full gauge invariant set of electroweak radiative corrections with the computer algebra approach. However, this full-precision approach is both time- and resource-consuming, and may not be necessary in all cases. We show that our approximate equations, obtained on paper, are in a very good the agreement with the full numerical results obtained with computer algebra in the low- and high-energy regions, and may be able to provide sufficient precision while being much user-friendly.

\begin{acknowledgements}
Many thanks to Michael Roney for enlightening discussion regarding the new physics search at Belle-II. This work was supported by
the Natural Sciences and Engineering Research
Council of Canada, the Harrison McCain Foundation which funded Dr. Zykunov's visit to Acadia University.
\end{acknowledgements}


\begin{thebibliography}{10}
	\providecommand{\url}[1]{{#1}}
	\providecommand{\urlprefix}{URL }
	\expandafter\ifx\csname urlstyle\endcsname\relax
	\providecommand{\doi}[1]{DOI \discretionary{}{}{}#1}\else
	\providecommand{\doi}{DOI \discretionary{}{}{}\begingroup
		\urlstyle{rm}\Url}\fi
\bibitem {LRP} Canadian Subatomic Physics Long Range Plan, 2017-2021, \url{http://www.subatomicphysics.ca} (2016).

\bibitem {hollik} W. Hollik,  Fortschr. Phys.  {\bf 38},  165   (1990). 

\bibitem {BH82} M. Bohm, W. Hollik, Nucl. Phys. B. {\bf 204}, 45 (1982).

\bibitem {BH84} M. Bohm, W. Hollik, Z. Phys. C.      {\bf 23},   31 (1984).

\bibitem {LEPTOP}  V.A. Novikov, L.B. Okun, M.I. Vysotsky, Nucl. Phys. B.    {\bf 397},   35 (1993).

\bibitem {TOPAZ96} G.  Montagna {\it et al.}, Comput. Phys. Commun.     {\bf 117},  278 (1999).

\bibitem {ZF91} D.  Bardin {\it et al.}, Comput. Phys. Commun.      {\bf 133},   229 (2001).

\bibitem {grup-bar2} D.Yu. Bardin, S.  Rieman, T. Rieman, Z. Phys.     {\bf 32},  121  (1986).

\bibitem {KK} S. Jadach, B.F.L. Ward and Z. Was, Phys. Rev. D {\bf 63}, 113009  (2001).

\bibitem {sanc-eeff} A. Andonov {\it et al.}, Phys. Part. Nucl. {\bf 34}, 1125 (2003).	

	\bibitem{Hahn:2000kx}
	T.~Hahn, Comput.Phys.Commun. \textbf{140}, 418 (2001).
	
	\bibitem{Hahn:1998yk}
	T.~Hahn, M.~Perez-Victoria, Comput.Phys.Commun. \textbf{118}, 153 (1999).
	
	\bibitem{Vermaseren:2000nd}
	J.~Vermaseren,   (2000).
	\newblock \urlprefix\url{http://arxiv.org/abs/math-ph/0010025}
	
	\bibitem{Aleksejevs:2007pd}
	A.~Aleksejevs, S.~Barkanova, P.G. Blunden, J. Phys. \textbf{G36}, 045101
	(2009).
	
	\bibitem{Aleksejevs:2010nf}
	A.~Aleksejevs, S.~Barkanova, A.~Ilyichev, Y.~Kolomensky, V.~Zykunov,
	Phys.Part.Nucl. \textbf{44}, 161 (2013).
	
	\bibitem{Aleksejevs:2012zz}
	A.~Aleksejevs, S.~Barkanova, V.~Zykunov, Phys.Atom.Nucl. \textbf{75}, 209
	(2012).
	
	\bibitem{Aleksejevs:2010ub}
	A.~Aleksejevs, S.~Barkanova, A.~Ilyichev, V.~Zykunov, Phys. Rev. \textbf{D82},
	093013 (2010).
	
	\bibitem{Aleksejevs:2016whx}
	A.~Aleksejevs, S.~Wu, S.~Barkanova, Y.~Bystritskiy, V.~Zykunov,   (2016)
	
	\bibitem{Aleksejevs:2016tjd}
	A.G. Aleksejevs, S.G. Barkanova, V.A. Zykunov, Phys. Atom. Nucl.
	\textbf{79}(1), 78 (2016).
	\newblock [Yad. Fiz.79,no.1,20(2016)]
	
	\bibitem{Bohm:1986rj}
	M.~Bohm, H.~Spiesberger, W.~Hollik, Fortsch. Phys. \textbf{34}, 687 (1986).
	
	\bibitem{Denner:1991kt}
	A.~Denner, Fortschr. Phys. \textbf{41}, 307 (1993).
	\newblock \urlprefix\url{http://arxiv.org/abs/arXiv:0709.1075}
	
	
	\bibitem{Zykunov:2005tc}
	V.A. Zykunov, Phys. Rev. \textbf{D75}, 073019 (2007).
	
	\bibitem{Zykunov:2006yb}
	V.A. Zykunov, Phys. Atom. Nucl. \textbf{69}, 1522 (2006).
	\newblock [Yad. Fiz.69,1557(2006)]
	
\end{thebibliography}

\end{document}